\documentclass[aps,prd,10pt,amsmath,amssymb,twocolumn,floatfix,nosuperscriptaddress,preprintnumbers,nofootinbib]{revtex4-2}

\input{header}

\usepackage{listings}
\usepackage{bm}
\usepackage{color}

\makeatletter
\def\l@subsection#1#2{}
\def\l@subsubsection#1#2{}
\makeatother

\begin{document}

\preprint{UCI-TR-2025-10}

\title{Characterizing Heavy Neutral Leptons: Measuring Parameters, Discriminating Majorana versus Dirac, and Using FASER2 as a Trigger for ATLAS}

\author{Jonathan L.~Feng}
\email{jlf@uci.edu}
\affiliation{Department of Physics and Astronomy, University of California, Irvine, CA 92697, USA}

\author{Alec Hewitt}
\email{ahewitt1@uci.edu}
\affiliation{Department of Physics and Astronomy, University of California, Irvine, CA 92697, USA}

\author{Daniel La Rocco}
\email{laroccod@uci.edu}
\affiliation{Department of Physics and Astronomy, University of California, Irvine, CA 92697, USA}

\author{Daniel Whiteson\vspace*{0.1in}}
\email{daniel@uci.edu}
\affiliation{Department of Physics and Astronomy, University of California, Irvine, CA 92697, USA}

\begin{abstract}
This work explores the potential of the proposed FASER2 experiment at the LHC to determine the properties of a discovered heavy neutral lepton (HNL), including its mass, couplings, and whether it is a Majorana or Dirac fermion.  We first consider a Majorana HNL with mass $m_N = 1.84\,\gev$ that is primarily produced through decays $D \to \mu N$ at the ATLAS interaction point.  Such HNLs may travel macroscopic distances in the far-forward direction and then decay, yielding approximately 8600 $N \to \mu \pi$ decays in FASER2 at the High-Luminosity LHC. With FASER2 measurements alone, the HNL's mass and couplings can be measured to fractional uncertainties of approximately 0.1\% and 3\% at 95\% CL, respectively, and the Dirac fermion hypothesis can be rejected at 99.8\% CL. We then consider a second, more difficult, case of a Majorana HNL with mass $m_N = 2.00\,\gev$, yielding only 80 $N \to \mu \pi$ decays in FASER2.  With FASER2 alone,  measurements of HNL properties are still possible, but somewhat less precise. However, by using FASER2 as a trigger for ATLAS and measuring the charge of the muon produced in association with the HNL at ATLAS to search for lepton number violation, one can precisely measure the HNL's properties and reject the Dirac fermion hypothesis at 99.7\% CL.  These results show that FASER2, sometimes in coordination with ATLAS, can precisely determine HNL properties, with far-reaching implications for our understanding of neutrino masses, baryogenesis, and the fundamental symmetries of nature. 
\end{abstract}

\maketitle

\tableofcontents

\section{Introduction}
\label{sec:intro}

The Standard Model (SM) of particle physics, although remarkably successful, fails to explain several key observations: the small, but nonzero, masses of neutrinos; the dominance of matter over antimatter in the universe, and the existence of dark matter. These shortcomings point toward the need for physics beyond the SM (BSM). 

A fairly minimal way to extend the SM is to add sterile neutrinos, fermionic gauge eigenstates with no SM gauge quantum numbers.  When these mix with the active neutrinos of the SM, the resulting mass eigenstates are heavy neutral leptons (HNLs).  HNLs, with the right properties and possibly supplemented by other states, may address all of the above-mentioned problems: the mixing of sterile and active neutrinos can generate the observed neutrino masses and mixings~\cite{Weinberg:1979sa,Chen:2011de,King:2015aea,deGouvea:2016qpx}, the HNLs may provide a viable mechanism for baryogenesis via leptogenesis~\cite{Fukugita:1986hr,Davidson:2008bu}, and HNLs may be cold or warm dark matter~\cite{Gunn:1978gr,Dodelson:1993je,Abazajian:2017tcc,Boyarsky:2018tvu}.

\begin{figure*}[tbph]
    \centering
\includegraphics[width=.95\linewidth]{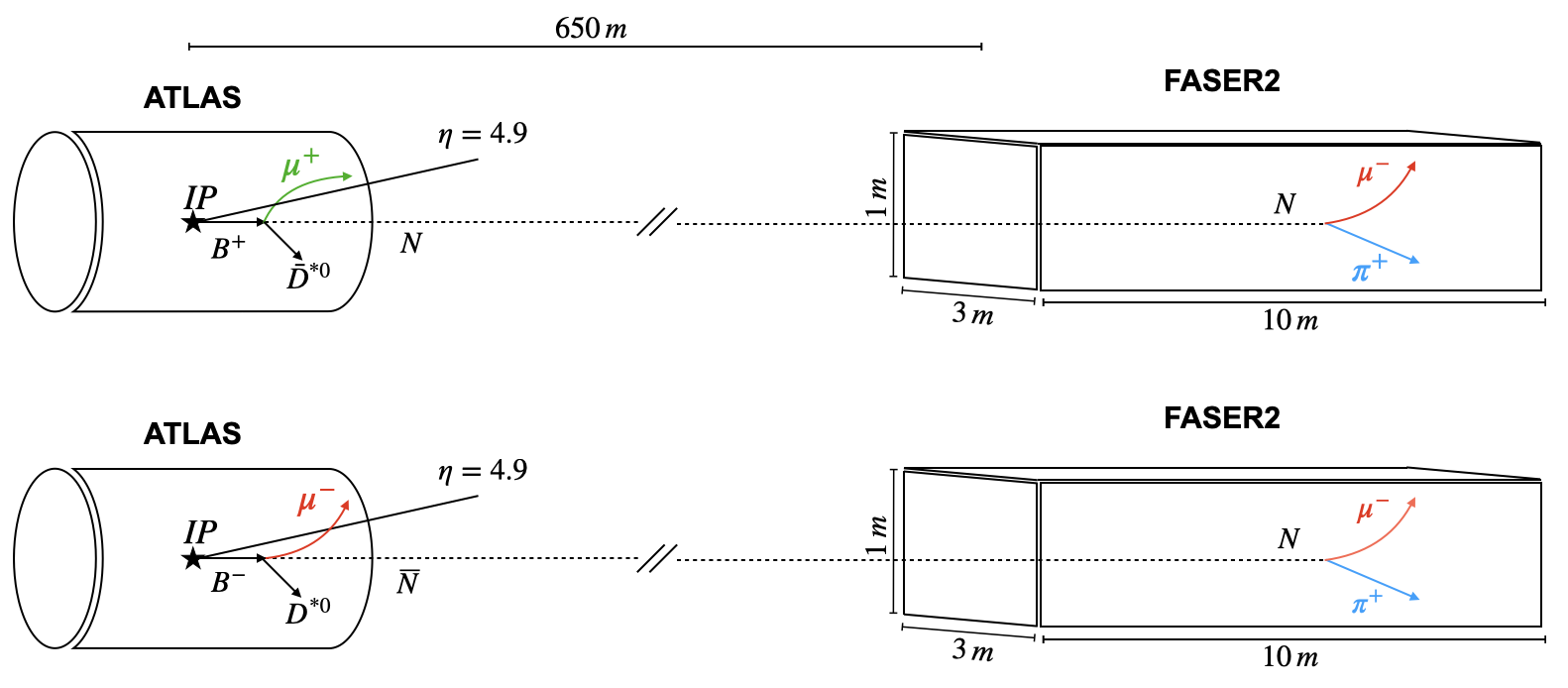}
\caption{Schematic diagrams showing example HNL events at ATLAS and FASER2~\cite{Salin:2927003}.  A charged $B$ meson decays through $B^{\mp} \to \overset{\scriptscriptstyle(-)}{D}{}^{*0}\mu^{\mp} N$ at the ATLAS IP.  The HNL then travels approximately 650 m and decays through $N \to \mu^{\mp} \pi^{\pm}$ in FASER2.  For Majorana HNLs, both the top (lepton-number conserving) and bottom (lepton-number-violating) processes are allowed. For Dirac HNLs, only the top process is allowed. } 
    \label{fig:faser-atlas}
\end{figure*}

Given these motivations, HNLs have been the target of experimental searches for many years~\cite{Bjorken:1972am,Shrock:1974nd,Shrock:1980vy,Shrock:1980ct,Shrock:1981wq}; for recent summaries of HNL searches and existing constraints, see Refs.~\cite{Boser:2019rta,Fernandez-Martinez:2023phj}. Once discovered, however, attention will immediately turn to measuring the HNL's masses and couplings and to determine if they are Majorana or Dirac fermions.  These measurements will be essential, not only to evaluate whether they have the right properties to solve the puzzles noted above, but also because their properties may point toward other particles, providing a target for future colliders, facilities, and experiments. 

In this work, we consider the possibility that HNLs are discovered in forward detectors at the Large Hadron Collider (LHC).  Specifically, we consider a successor to the currently-running FASER experiment~\cite{Feng:2017uoz,FASER:2022hcn}, FASER2~\cite{Salin:2927003}, which will be located 650 m downstream of the ATLAS interaction point (IP), along the beam collision axis or line-of-sight (LOS), in the proposed Forward Physics Facility~\cite{Feng:2022inv,Adhikary:2024nlv,FPFWorkingGroups:2025rsc}.  At the LHC, HNLs with weak couplings and masses in the MeV to few GeV range can be produced in very large numbers in a highly collimated beam along the LOS, travel hundreds of meters, and then decay to visible SM particles.  FASER2 is therefore ideally positioned to discover HNLs in regions of parameter space that are currently unconstrained by particle experiments and cosmology, and the possibility of detecting HNLs in far-forward experiments at the LHC has been the subject of numerous studies~\cite{Kling:2018wct,Helo:2018qej,FASER:2018eoc,Cottin:2021lzz,Mao:2023zzk,Dreiner:2023gir,Feng:2024zfe,Wang:2024mrc,Bernal:2024kqx}.  

In this paper, we assume that an HNL has been discovered, and we explore the potential of FASER2 to constrain its mass and couplings and determine its spinor nature, that is, if it is a Majorana or Dirac fermion.  We begin by considering a model where a Majorana HNL has mass 1.84 GeV and couples to muons.  In this model, the HNL is dominantly produced in decays $D_s^{\pm} \to \mu^{\pm} N$ at the ATLAS IP and then travels a macroscopic distance before decaying; see \cref{fig:faser-atlas}.  This model is currently allowed by all constraints, but it predicts remarkable event rates for FASER2: at the High-Luminosity LHC (HL-LHC) with its expected integrated luminosity of $3~\text{\ab}^{-1}$, $\sim 10^9$ HNLs are produced by ATLAS, $\sim 10^5$ of these decay in FASER2, and approximately 8600 of these are $N \to \mu \pi$ decays.  By measuring the energy and momentum of the HNL decay products at FASER2 using only this fully-visible decay mode, we show that the HNL mass and coupling can be constrained to fractional uncertainties of $0.1\%$ and $3\%$ at 95\% confidence level (CL), respectively, and the possibility that the HNL is a Dirac fermion can be disfavored at 99.8\% CL ($2.9 \sigma$), provided current uncertainties in forward hadron fluxes can be significantly reduced by future far-forward neutrino measurements, as is expected~\cite{Kling:2023tgr} and will be discussed below.  

Of course, HNLs may be discovered with significantly fewer events.  We therefore then consider a Majorana HNL with mass 2 GeV, for which only roughly 80 $N \to \mu \pi$ decays in FASER2 are expected at the HL-LHC.  For this model, standalone FASER2 measurements are able to constrain the HNL mass and coupling to fractional uncertainties of $1\%$ and $25\%$ at 95\% CL, respectively, and the possibility that the HNL is a Dirac fermion can be disfavored at 98.2\% CL ($2.1 \sigma$). 

In this second model, however, we also note that a sizable fraction of FASER2 HNL events are accompanied by muons that can be observed in ATLAS.  We, therefore, also consider the possibility of using FASER2 as a trigger for ATLAS and combining the resulting data from FASER2 and ATLAS.  An HNL travels from the ATLAS IP to FASER2 in $2.2~\mu\text{s}$, and a signal may be sent back to ATLAS through an optical cable in $3.2~\mu\text{s}$.  Including the time it takes to identify a signal at FASER2 as being interesting, the resulting roundtrip time is (just) within the ATLAS Level-0 (L0) trigger latency window of $6~\mu\text{s}$.  

As shown in \cref{fig:faser-atlas}, if the muon produced in association with the HNL can be recorded in ATLAS, its properties may be used to sharpen FASER2's measurements.   In particular, by measuring the muon's charge, one can determine if the complete HNL event is lepton-number-conserving (LNC) (\cref{fig:faser-atlas}, top) or lepton-number-violating (LNV) (\cref{fig:faser-atlas}, bottom), and distinguish between Majorana and Dirac HNLs.  In the absence of background, events may be classified as LNC or LNV on an event-by-event basis, and a single unambiguous LNV event may provide conclusive evidence that the HNL is a Majorana fermion.  Unfortunately, the muons produced in association with HNLs at ATLAS are very far-forward, and there are daunting backgrounds in this region, especially given pileup at the HL-LHC.  Despite this, we find that, assuming detector upgrades to ensure muon detection up to pseudorapidities of $|\eta| < 4.9$, suitable cuts allow $\text{FASER2} + \text{ATLAS}$ to differentiate between Majorana and Dirac HNLs and reject the Dirac fermion hypothesis at 99.7\% CL.

Several strategies have been proposed in the literature to distinguish Majorana from Dirac HNLs by exploiting differences between LNC and LNV processes~\cite{Tastet:2019nqj,Mikulenko:2023ezx,Bolton:2022tds,Drewes:2022rsk}. Previous work has also explored the use of the far-forward detector SND@LHC to trigger ATLAS~\cite{SNDLHC:2025qtx}, and a recent study has shown that FASER can distinguish Majorana and Dirac HNLs through angular measurements in the $e^+e^-$ final states~\cite{A:2025gpb}.  In contrast, our work uses energy and invariant mass distributions, and is the first to utilize a far-forward detector as a trigger for ATLAS to probe the spinor nature of HNLs.  

We note also that, very recently, the first candidate ``tagged'' neutrino event~\cite{Pontecorvo:1979zh} has been reported by the NA62 Collaboration~\cite{NA62:2024xzj}, where the neutrino's production in the decay $K^+ \to \mu^+ \nu_{\mu}$ is observed, and then its subsequent charged-current (CC) interaction in a liquid krypton calorimeter is also detected. In this study, we extend the detection of tagged neutrinos in the SM to the detection of tagged HNLs in BSM models.  This work demonstrates that, with ``cradle to grave'' observations that track HNLs throughout their entire lifetime, one can precisely determine the HNL's properties, including if it is a Majorana fermion.  Such a determination will have far-reaching consequences for our understanding of neutrino masses, baryogenesis, and the fundamental symmetries of nature. 

This study is organized as follows.  In \cref{sec:Model}, we describe the HNL model, including the relevant Lagrangian, mixing parameters, and the resulting interactions. In \cref{sec:Simulating HNL signals at FASER2}, we explain our experimental setup and simulation, and in \cref{sec:statistics} we detail the statistical methods we will use. In \cref{sec:Measuring model paremeters with FASER2}, we assess how measurements at FASER2 alone can measure HNL masses and couplings. In \cref{sec:Discrimination on the basis of lifetime}, we show that these same measurements may be used to determine if an HNL is a Majorana or Dirac fermion.  In \cref{sec:FASER2 as a trigger for ATLAS}, we consider the more challenging HNL model, and investigate the potential of using FASER2 as a trigger for ATLAS, showing how the charge of the lepton produced in association with the HNL can be used to improve model discrimination. Finally, in \cref{sec:Conclusion} we summarize our findings.

\section{The HNL Model}
\label{sec:Model}

We consider an extension of the SM with sterile neutrinos. The Lagrangian is given by
\begin{align}
\label{eq:lagrangian}
    \mathcal{L} \supset &\ i \bar{N}'_i \slashed{\partial} P_R N'_i  
    - \frac{m_{\alpha \beta}}{2} \overline{\nu_{\alpha}^c} P_L \nu_{\beta}
    - y_{\alpha  i}  \, \overline{L}_\alpha \,\tilde{\phi} \, P_R N^\prime_i  \nonumber \\
  &  - \frac{M_{ij}}{2} \, \overline{{N^\prime_i}^c} \, P_R N^\prime_j + \text{h.c.} \ ,
\end{align}
where all of the fermions are four-component spinors, the $N'_i$ are gauge-singlet sterile neutrinos, $L_{\alpha} = (\nu_{\alpha}, l_{\alpha})^T$ are SM lepton doublets, $\tilde{\phi} = i \sigma_2 \phi^*$ is the conjugated Higgs doublet, $m_{\alpha\beta}$ are Majorana masses for active neutrinos, $M_{ij}$ are Majorana masses for the sterile neutrinos, and $y_{\alpha i}$ are Yukawa couplings. 

After electroweak symmetry breaking, the Yukawa interaction generates a Dirac mass $m_D$, and the neutrino mass terms can be arranged into the matrix
\begin{equation}
    \mathcal{M} = \begin{pmatrix}
        m & m_D \\
        m_D^T & M
    \end{pmatrix} ,
\end{equation}
which acts on the column vector $(P_R \nu_\alpha^c,\, P_R N')^T$. Diagonalizing this mass matrix yields mass eigenstates, which are the physical neutrino states.

For simplicity, we focus on a minimal scenario involving a single HNL. In what follows, $N$ represents either a Dirac spinor, composed of 2 independent Weyl spinors, or a Majorana spinor, composed of a single Weyl spinor. The flavor eigenstates can be written as
\begin{equation}
    P_L \nu_\alpha = \sum_{i = 1}^{3} V_{\alpha i} \, P_L \nu_i + U_\alpha P_L N^c ,
\end{equation}
where $U_\alpha$ parametrizes the active–sterile mixing. By substituting $\nu_\alpha$ into the SM Lagrangian, one finds that the HNL couples to the weak gauge bosons through the charged-current (CC) and neutral-current (NC) interactions
\begin{equation}
\begin{split}
\mathcal{L}^{\text{CC}} &= - \frac{g}{\sqrt{2}} U^*_\alpha  W_\mu^+  \,\overline{N^c} \, \gamma^\mu P_L l_\alpha + \text{h.c.} \\
\mathcal{L}^{\text{NC}} &=  - \frac{g}{2\cos\theta_W} U^*_\alpha Z_\mu  \overline{N^c}\, \gamma^\mu P_L \nu_\alpha+ \text{h.c.}
    \label{eq:CCNC}
\end{split}
\end{equation}
The interactions of \cref{eq:CCNC} allow hadrons to decay to HNLs through CC interactions, as well as HNLs to decay back into SM particles through NC and CC interactions. 

\begin{table*}[tbph]
\begin{center}
\begin{tabular}{|c|c|c|c|c|c|c|}
\hline
\multicolumn{2}{|c}{\textbf{}} & \multicolumn{2}{|c}{\textbf{Model 1}} & \multicolumn{3}{|c|}{\textbf{Model 2}} \\
\hline
\multicolumn{2}{|c}{HNL Model Parameters} & \multicolumn{2}{|c}{ \ $m_N = 1.84~\gev$,  $U_{\mu} = 0.0036$, Majorana \ } & \multicolumn{3}{|c|}{\ $m_N = 2.00~\gev$,  $U_{\mu} = 0.002$, Majorana \ } \\
\hline
\multicolumn{2}{|c}{$c \tau_N$} & \multicolumn{2}{|c}{2.03 m} & \multicolumn{3}{|c|}{4.33 m} \\
\hline
 \multicolumn{2}{|c}{$\langle E_N \rangle_{\rm{FASER2}}$} & \multicolumn{2}{|c}{820 GeV} & \multicolumn{3}{|c|}{828 GeV} \\
\hline

 \multicolumn{2}{|c}{$\langle E_N \rangle_{\rm{FASER2}+\rm{ATLAS}}$} & \multicolumn{2}{|c}{458 GeV} & \multicolumn{3}{|c|}{452 GeV} \\
\hline

\multicolumn{1}{|c}{} & \multicolumn{1}{|c}{} & \multicolumn{1}{|c}{} & \multicolumn{1}{|c}{FASER2} & \multicolumn{1}{|c}{} & \multicolumn{1}{|c}{FASER2} & \multicolumn{1}{|c|}{\ FASER2+ATLAS \ } \\
\hline
\multicolumn{1}{|c}{Production Modes} & \multicolumn{1}{|c}{$N_{\rm{events}}$} & \multicolumn{1}{|c}{$D_s \to \mu N$} & \multicolumn{1}{|c}{119,000} & \multicolumn{1}{|c}{$B \to D^{*0} \mu N$} & \multicolumn{1}{|c}{447} & \multicolumn{1}{|c|}{56.1} \\
\multicolumn{1}{|c}{} & \multicolumn{1}{|c}{at HL-LHC} & \multicolumn{1}{|c}{$B \to D^{*0} \mu N$} & \multicolumn{1}{|c}{1670} &  \multicolumn{1}{|c}{$B^0 \to D^* \mu N$} & \multicolumn{1}{|c}{405} & \multicolumn{1}{|c|}{50.7} \\
\multicolumn{1}{|c}{} & \multicolumn{1}{|c}{(in $3~\text{ab}^{-1}$)} & \multicolumn{1}{|c}{$B^0 \to D^{*} \mu N$} & \multicolumn{1}{|c}{1570} & \multicolumn{1}{|c}{$B \to D^0 \mu N$} & \multicolumn{1}{|c}{199} & \multicolumn{1}{|c|}{26.4}  \\
\multicolumn{1}{|c}{} & \multicolumn{1}{|c}{} & \multicolumn{1}{|c}{$B^0_s \to D^*_s \mu N$} & \multicolumn{1}{|c}{333} & \multicolumn{1}{|c}{$B^0 \to D \mu N$} & \multicolumn{1}{|c}{183} & \multicolumn{1}{|c|}{22.4} \\

\hline

\multicolumn{1}{|c}{Decay Modes} & \multicolumn{1}{|c}{Branching} & \multicolumn{1}{|c|}{$\mu \rho$} & \multicolumn{1}{c}{0.187} & \multicolumn{1}{|c}{$\mu \rho$} & \multicolumn{2}{|c|}{0.162} \\
\multicolumn{1}{|c}{} & \multicolumn{1}{|c}{Fraction} & \multicolumn{1}{|c|}{$\nu \nu \nu$} & \multicolumn{1}{c}{0.129} & \multicolumn{1}{|c}{$\nu \nu \nu$} & \multicolumn{2}{|c|}{0.128} \\
\multicolumn{1}{|c}{} & \multicolumn{1}{|c}{}& \multicolumn{1}{|c|}{$\nu e \mu$} & \multicolumn{1}{c}{0.125} & \multicolumn{1}{|c}{$\nu e \mu$} & \multicolumn{2}{|c|}{0.125} \\
\multicolumn{1}{|c}{} & \multicolumn{1}{|c}{} & \multicolumn{1}{|c|}{$\nu \mu \mu$} & \multicolumn{1}{c}{0.072} & \multicolumn{1}{|c}{$\nu \mu \mu$} & \multicolumn{2}{|c|}{0.073} \\

\multicolumn{1}{|c}{} & \multicolumn{1}{|c}{}& \multicolumn{1}{|c|}{$\bm{\mu \pi}$} & \multicolumn{1}{c}{\textbf{0.071}} & \multicolumn{1}{|c}{$\bm{\mu \pi}$} & \multicolumn{2}{|c|}{\textbf{0.060}} \\
\multicolumn{1}{|c}{} & \multicolumn{1}{|c}{}& \multicolumn{1}{|c|}{$\nu \pi^0$} & \multicolumn{1}{c}{0.038} & \multicolumn{1}{|c}{$\nu \pi^0$} & \multicolumn{2}{|c|}{0.032}  \\
\multicolumn{1}{|c}{} & \multicolumn{1}{|c}{} &  \multicolumn{1}{|c|}{Other} & \multicolumn{1}{c}{0.378} & \multicolumn{1}{|c}{Other} & \multicolumn{2}{|c|}{0.419} \\
\hline
\multicolumn{2}{|c }{\ Total FASER2 $N \to \mu \pi$ Events} \ & & {8610} & 
\multicolumn{1}{|c}{} &
\multicolumn{1}{|c}{81.8} & \multicolumn{1}{|c|}{10.6} \\
\hline
\end{tabular}
\caption{Summary of the two benchmark HNL models considered in this study, including the HNL mass $m_N$, active-sterile mixing parameter $U_\mu$, and decay length $c \tau_N$.  $\langle E_N \rangle_{\rm{FASER2}}$ is the average energy of HNLs that decay in FASER2, while $\langle E_N \rangle_{\rm{FASER2}+\rm{ATLAS}}$ is the average energy of HNLs that decay in FASER2 and are accompanied by a signal muon that passes through ATLAS.  The dominant production modes are listed, along with the number of HNLs they produce that can be detected at FASER2 and FASER2+ATLAS, assuming the baseline HL-LHC integrated luminosity of $3~\text{ab}^{-1}$. The dominant decay modes and their branching fractions are also listed.  All production and decay modes include both charge-conjugate processes.  The fully-reconstructible decay mode $N \to \mu \pi$ used in this study is highlighted.  Model 1 represents models with a large number of decays in FASER2, while Model 2 has far fewer decays in FASER, but predicts that a sizable fraction of these decays are accompanied by signal muons that pass through ATLAS. }
\label{tab:model_summary}
\end{center}
\end{table*}

For this work, we consider HNLs that only couple to muons, so $U_{\mu} \neq 0$, but $U_{e} = U_{\tau} = 0$.  The analysis could be extended to HNLs with electron or tau couplings, with the typical caveat that the experimental signals may be more challenging to detect, particularly for tau couplings.  For the muon case we consider here, the dominant production modes include $P \to \mu N$, $P \to P' \mu N$, and $P \to V \mu N$, where $P$ and $P'$ denote pseudoscalar mesons, and $V$ denotes a vector meson, as well as production from $\tau$ decays $\tau \to \mu \nu N$. Possible decay channels of the HNL include $N \to H^0 \nu_{\mu}$, $N \to H \mu$, $N \to \ell \ell \nu_{\mu}$, $N \to \ell \nu_{\ell} \mu$, and $N \to \bar{\nu}_{\ell} \nu_{\ell} \nu_{\mu}$, where $H$ denotes either a pseudoscalar or a vector meson, and $\ell = e,\,\mu,\,\tau$.  For a full discussion of HNL production and decay modes, see Ref.~\cite{Feng:2024zfe}. 

In this study, we examine a Majorana HNL with mass $m_N$ and coupling $U_\mu$, and test how well the signal can be described by a Dirac HNL with parameters $m_N'$ and $U_\mu'$. Because LNV decay modes are allowed for Majorana HNLs, but forbidden for Dirac HNLs, the lifetimes of Majorana HNLs are related to the lifetimes of their Dirac counterparts by 
\begin{align}
\tau_{\text{Majorana}} = \frac{1}{\Gamma_{\text{LNC}} + \Gamma_{\text{LNV}}} = \frac{1}{2} \tau_{\text{Dirac}} \ .
\end{align}

As discussed in \cref{sec:intro}, we will consider two representative HNL models.  A summary of their relevant properties is given in \cref{tab:model_summary}. In Model 1 the HNL mass lies just below the $D_s$ threshold, allowing prolific production of HNLs from forward $D_s \to \mu N$ decays at the LHC. In contrast, in Model 2, the HNL mass lies above the $D_s$ threshold, and HNLs can only be produced in $B$ decays, resulting in lower overall yields, but the muons produced in association with the HNLs are produced at lower $\eta$, allowing them to be detected in ATLAS. Both benchmark points lie in currently unconstrained regions of parameter space and are within reach of FASER2, offering promising opportunities to probe the Majorana vs.~Dirac nature of HNLs through event rates, kinematics, and decay signatures.  In this work, we focus solely on the $\mu\pi$ decay mode, where the decay products are fully visible and the HNL four-momentum is, in principle, reconstructible.  Other decay modes, notably $\mu\rho$, may also be useful and can be expected to improve the results derived here.

\section{Simulating HNL Signals at FASER2}
\label{sec:Simulating HNL signals at FASER2}

FASER2~\cite{Salin:2927003,Feng:2022inv,Adhikary:2024nlv} is a proposed upgrade of the FASER experiment.  As illustrated in \cref{fig:faser-atlas}, we assume that FASER2 is centered on the LOS, and the front of the FASER2 detector is $L = 650~\text{m}$ from the ATLAS IP. The FASER2 decay volume is 10 m long, with a rectangular cross section that is 1 m high and 3 m wide.

The detector includes several key components critical to reconstructing HNL decay events. An electromagnetic calorimeter stops electrons and photons and measures their energies. A hadronic calorimeter measures the energy of charged pions and distinguishes them from muons. A tracking system measures muon momenta and distinguishes $\mu^-$ from $\mu^+$ through their curvature in a magnetic field. Estimates of the uncertainties of these measurements for the current FASER2 design~\cite{FASER2resolutions} are
\begin{align}
\frac{\sigma_{p_{\mu}}}{p_{\mu}} & = 3.3 \times 10^{-5} \, \left[ p_{\mu} / \text{GeV} \right] \\
\frac{\sigma_{E_{\pi^{\pm}}}}{E_{\pi^{\pm}}} & = 
\frac{35\%}{\sqrt{E_{\pi^{\pm}} / \text{GeV}}} \oplus 0.5\% \\
\sigma_{\theta_{\mu \pi}} & = 250~\mu\text{rad} \ . 
\label{eq:exptresolution}
\end{align}

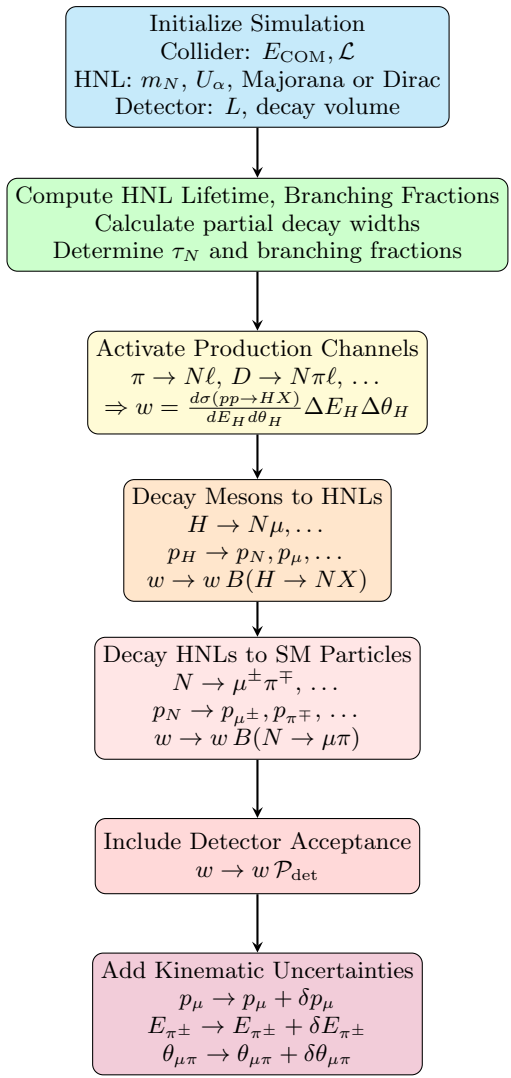
\begin{figure}
\begin{center}
\tikzstyle{startstop} = [rectangle, rounded corners, minimum width=3cm, minimum height=1cm, text centered, draw=black, fill=cyan!20, align=center]
\tikzstyle{lifetime} = [rectangle, rounded corners, minimum width=3cm, minimum height=1cm, text centered, draw=black, fill=green!20, align=center]
\tikzstyle{production} = [rectangle, rounded corners, minimum width=3cm, minimum height=1cm, text centered, draw=black, fill=yellow!20, align=center]
\tikzstyle{pN} = [rectangle, rounded corners, minimum width=3cm, minimum height=1cm, text centered, draw=black, fill=orange!20, align=center]
\tikzstyle{pdaught} = [rectangle, rounded corners, minimum width=3cm, minimum height=1cm, text centered, draw=black, fill=red!10, align=center]
\tikzstyle{smear} = [rectangle, rounded corners, minimum width=3cm, minimum height=1cm, text centered, draw=black, fill=purple!20, align=center]
\tikzstyle{decay} = [rectangle, rounded corners, minimum width=3cm, minimum height=1cm, text centered, draw=black, fill=red!15, align=center]

\tikzstyle{arrow} = [thick, ->, >=stealth]

\begin{tikzpicture}[node distance=2.1cm]

\node (initialize) [startstop] 
{Initialize Simulation\\
Collider: $E_{\text{COM}}, \mathcal{L}$\\
HNL: $m_N$, $U_\alpha$, Majorana or Dirac\\
Detector: $L$, decay volume};

\node (lifetime) [lifetime, below of=initialize] 
{Compute HNL Lifetime, Branching Fractions\\
Calculate partial decay widths\\
Determine $\tau_N$ and branching fractions};

\node (production) [production, below of=lifetime] 
{Activate Production Channels\\
$\pi \to N \ell$, $D \to N \pi \ell$, $\dots$\\
$\Rightarrow w = \frac{d\sigma(pp \to H X)}{d E_H d\theta_H} \Delta E_H \Delta \theta_H$};

\node (pN) [pN, below of=production] 
{Decay Mesons to HNLs\\
$H \to N \mu, \dots$\\
$p_H \to  p_N, p_{\mu}, \dots$\\
$w \to w \, B(H \to N X)$};

\node (pdaught) [pdaught, below of=pN] 
{Decay HNLs to SM Particles\\
$N \to \mu^{\pm} \pi^{\mp}$, \ldots\\
$p_N \to p_{\mu^{\pm}}, p_{\pi^{\mp}}$, \ldots\\
$w \to w \, B(N \to \mu \pi)$};

\node (decay) [decay, below of=pdaught] 
{Include Detector Acceptance\\
$w \to w \, \mathcal{P}_{\text{det}}$};

\node (smear) [smear, below of=decay]{Add Kinematic Uncertainties\\
$p_{\mu} \to p_{\mu} + \delta p_{\mu}$\\
$E_{\pi^{\pm}} \to E_{\pi^{\pm}} + \delta E_{\pi^{\pm}}$\\
$\theta_{\mu \pi} \to \theta_{\mu \pi} + \delta \theta_{\mu \pi}$};

\draw [arrow] (initialize) -- (lifetime);
\draw [arrow] (lifetime) -- (production);
\draw [arrow] (production) -- (pN);
\draw [arrow] (pN) -- (pdaught);
\draw [arrow] (pdaught) -- (decay);
\draw [arrow] (decay) -- (smear);
\end{tikzpicture}
\end{center}
\caption{Schematic diagram illustrating the major steps in simulating HNL events in the ATLAS and FASER2 detectors. The parameter $w$ is the event weight; it starts with the number of hadrons produced for a particular range of energy and angle, and then is reduced by branching fractions and detector acceptance to determine the number of HNL events detected in FASER2.}
\label{fig:FORESEE}
\end{figure}

In addition to these uncertainties, there is a systematic uncertainty on the HNL flux from uncertainties in the flux of forward hadron production at the LHC. We estimate this uncertainty by using the event generators in \texttt{FORESEE}~\cite{Kling:2021fwx} that yield the maximal and minimal number of HNL events. For the minimum, we used \texttt{QGSJET 2.04}~\cite{Ostapchenko:2010vb} for light hadrons and \texttt{NLO-P8-Min} for heavy hadrons. For the maximum, we used \texttt{SIBYLL 2.3d}~\cite{Riehn:2019jet} for light hadrons and \texttt{NLO-P8-Max} for heavy hadrons. The heavy hadron generators are based on \texttt{Pythia8}~\cite{Bierlich:2022pfr}, where the underlying assumptions are detailed in Ref.~\cite{Buonocore:2023kna}. The range of event rates from these generators corresponds to an overall flux uncertainty of approximately $\pm 60\%$.

To perform realistic event simulations, we use the \texttt{FORESEE} (FORward Experiment SEnsitivity Estimator) framework~\cite{Kling:2021fwx} for event generation and \texttt{HNLCalc}~\cite{Feng:2024zfe,HNLCalc2024} to extend \texttt{FORESEE} to simulate HNL events.  A diagram illustrating the simulation workflow is shown in~\cref{fig:FORESEE}.  In \texttt{FORESEE}, one specifies the $pp$ center-of-mass energy $E_{\text{COM}} = 14~\tev$ and integrated luminosity $\mathcal{L} = 3~\ab^{-1}$ for the HL-LHC, and chooses one of several state-of-the-art Monte-Carlo (MC) generators to generate the forward meson production cross sections and spectra, as discussed above.  One then further specifies the HNLs mass $m_N$, mixing angles $U_\alpha$, and spinor nature (Majorana or Dirac), and \texttt{HNLCalc} then computes the production rate for HNLs in meson decays, and the HNL decay branching fractions and lifetime.  Finally, one must specify the detector's distance from the ATLAS IP $L$ and its decay volume geometry.  

With all of this information, \texttt{FORESEE} then simulates the production of HNLs in meson decays, propagates them to the FASER2 detector, and decays them in the specified decay volume. As a last step, we crudely simulate the detector's response by adding a Gaussian smearing of the relevant kinematic quantities with the resolutions given in \cref{eq:exptresolution}. The output is an estimate of the number of HNL events detected in FASER2, along with all relevant kinematic distributions.  

\begin{figure*}
\centering 
\includegraphics[width=0.49\linewidth]{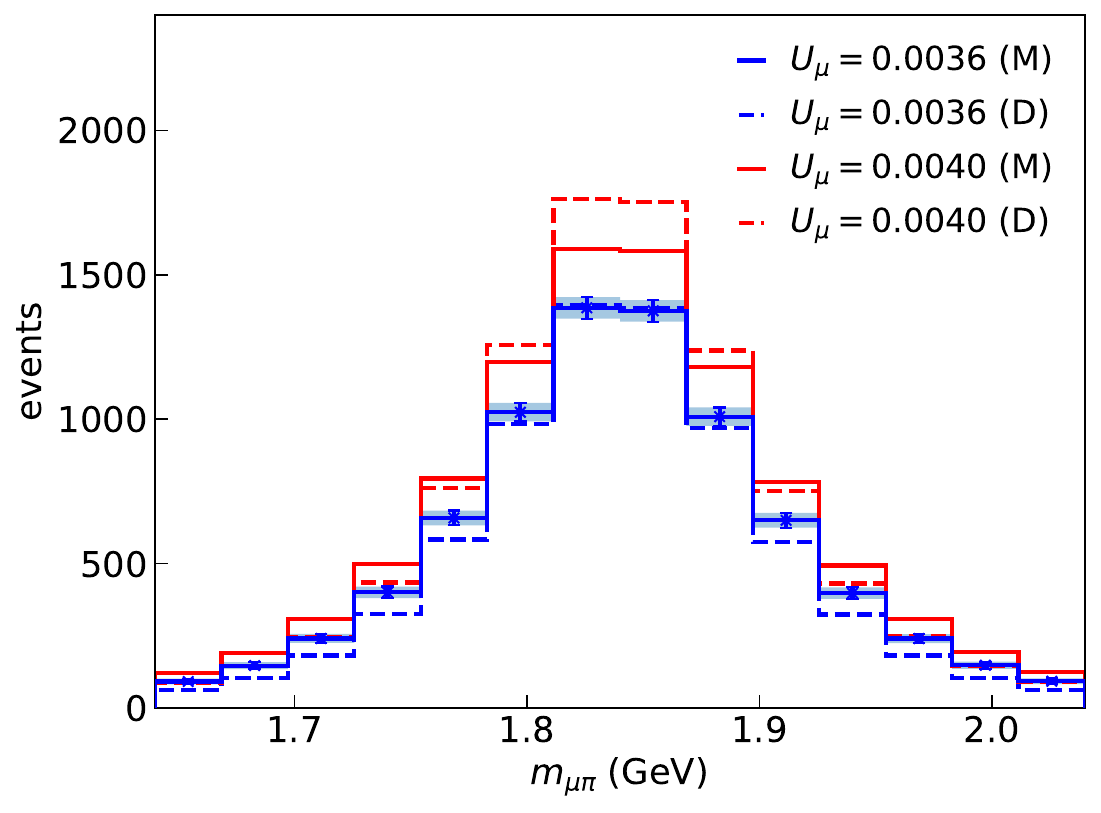}
\hfill  
\includegraphics[width=0.49\linewidth]{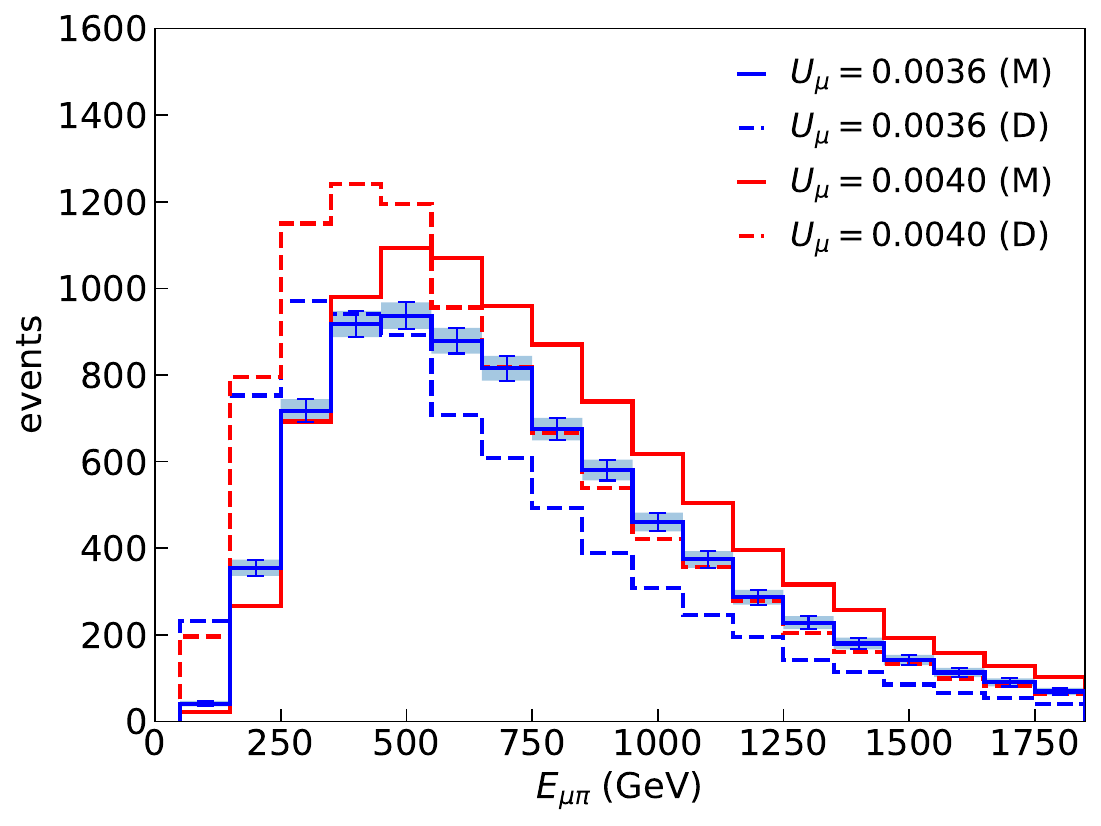}
\caption{Kinematic distributions for HNLs that decay through $N \to \mu^{\pm} \pi^{\mp}$ in FASER2.  The HNLs are produced in meson decays at the HL-LHC with $E_{\text{COM}} = 14~\tev$ and an integrated luminosity of $3~\text{ab}^{-1}$. Results are shown for Model 1, a Majorana HNL with $m_N = 1.84~\gev$ and $U_\mu = 0.0036$, as well as for variations from this model with Dirac HNLs, and with $U_\mu = 0.0040$, as indicated. The shaded error bars illustrate the envelope of expected outcomes in each bin, given by $\sqrt{\mu_i}$, where $\mu_i$ is the number of events in bin $i$. } 
\label{fig: 1d hists}
\end{figure*}

In \cref{fig: 1d hists} we show results for Model 1 of the simulation for one-dimensional kinematic distributions reconstructed from HNLs that decay via $N \to \mu^{\pm} \pi^{\mp}$ in FASER2.  The invariant mass is $m_{\mu \pi} = \left[ (p_{\mu} + p_{\pi})^2 \right]^{1/2}$, and the total energy is $E_{\mu \pi} = E_{\mu} + E_{\pi}$.  Although the full results are two-dimensional histograms, we display the one-dimensional projections for clarity.  Kinematic smearing has been included, as described above. All available meson production channels are included in the simulation.  For Model 1, approximately 8600 $N \to \mu^{\pm} \pi^{\mp}$ events are expected at the HL-LHC.  

Results are shown for Model 1, a Majorana HNL with $m_N = 1.84~\gev$ and $U_\mu = 0.0036$, as well as for variations from this model. The overall event rate scales roughly as $U_\mu^4$, and so there are significantly more events for $U_\mu = 0.004$ than for $U_\mu = 0.0036$.  The $m_{\mu \pi}$ distribution is centered at $m_N$ in all cases, as expected. However, the peak of the energy distribution shifts.  A larger $U_\mu$ implies a shorter HNL lifetime, requiring the HNLs to have higher energies to reach FASER2 before decaying. Similarly, for fixed $U_\mu$, switching from a Majorana to a Dirac HNL eliminates half of the decay modes, which increases the lifetime, leading to softer energy distributions.    

As can be seen in \cref{fig: 1d hists}, changing the HNL from a Majorana to a Dirac fermion or changing the coupling by a factor of 10\% leads to statistically significant changes in the number of events in the bins of these one-dimensional distributions, suggesting that these properties of HNLs can be precisely determined by FASER2 measurements alone.  In the following, we investigate whether these conclusions hold up under a more rigorous statistical analysis.

\section{Statistical Methods}
\label{sec:statistics}

\subsection{Profiled Likelihood}

To investigate the characterization capabilities of FASER2 in the event of a signal detection, we perform a binned maximum likelihood (ML) fit of the HNL parameters to an artificial dataset, referred to as an \textit{Asimov dataset}, $\vec{d}_A = (\mu_1, \ldots,\mu_N)$, where $\mu_i$ is the mean signal yield in bin $i$~\cite{Cowan:2010js}, allowing us to obtain the expected measurement sensitivity to the model parameters. We will separate our data into 2D bins over the invariant mass $m_{\mu\pi}$ and total energy $E_{\mu\pi}$, as we will discuss in more detail below in \cref{sec:Measuring model paremeters with FASER2}.

We model the likelihood of observing a given set of data, which we denote $\vec{d}$, as a Poisson distribution in each bin with Gaussian constraints from flux and MC uncertainties:
\begin{multline}
    L(\vec{d}\, |m_N , U_\mu \, ; \, \eta, \vec{\gamma}) \\
    = \prod_i f_P(d_i\,| \,\eta\gamma_i\mu_i)f_G(1|\eta,\sigma_\eta)f_G(1|\gamma_i,\sigma_i) \ ,
    \label{eq:likelihood}
\end{multline} 
where $\mu_i=\mu_i(m_N,U_\mu)$ is the expected bin count for a given $m_N$ and $U_\mu$; $\eta$ and $\vec{\gamma}$ are nuisance parameters that characterize the flux normalization and bin-by-bin uncertainties, respectively; and $f_P$ and $f_G$ are Poisson and Gaussian distributions, respectively. 

The parameter $\eta$ represents the systematic uncertainty in the HNL flux normalization. As a simplifying assumption, we have chosen to model this as a Gaussian distribution with a mean value of unity and a variance equal to $\sigma_\eta = 60\%$, in accordance with the size of the overall spread of fluxes predicted by various forward hadron generators, as discussed in \cref{sec:Simulating HNL signals at FASER2}. 

The parameters $\gamma_i$, on the other hand, represent the relative uncertainties in each bin. We incorporate the statistical uncertainty in  estimating $\mu_i$ that results from limited MC sampling. Additionally, to account for systematic uncertainties associated with the shape of the forward hadron flux, we add in quadrature a uniform fractional uncertainty, denoted $\sigma_\gamma$, which we estimate is currently roughly 10\%. In the large MC sample limit, we can model the bin-by-bin uncertainty as a Gaussian distribution with a mean value of unity and a variance of $\sigma_i = \frac{1}{\sqrt{n_i}} \oplus \sigma_
\gamma$, where $n_i$ is the number of MC samples in that bin.

\subsection{Estimating Model Parameters}

Using the Asimov dataset and the likelihood of \cref{eq:likelihood}, we perform a maximum likelihood estimation of the model parameters by minimizing the profiled likelihood
\begin{align}
    t(\vec{d} \,|\, m_N, U_\mu) := -2\ln L(\vec{d}\, | \, m_N , U_\mu \, ; \, \hat{\eta},\hat{\vec{\gamma}}) \ ,
\end{align}
where $\{\hat{\eta}\,(\vec{d}\,),\hat{\vec{\gamma}}\,(\vec{d}\,)\}$ are the ML estimates for the nuisance parameters that minimize $t$ for a given set of data $\vec{d}$. One can then assess the expected goodness-of-fit of the model parameters to the data by treating $t$ as a test-statistic and computing the median $p$ value, where 
\begin{equation}
    \label{eq:p good}
    p = \int_{t_\text{exp}}^{\infty}f_t(t'\,|\,\hat{m}_{N,A},\hat{U}_{\mu,A})\,dt' \ ,
\end{equation} 
and $f_t$ is the distribution of $t$, $t_\text{exp} = t(\vec{d}_A | \hat{m}_{N,A}, \hat{U}_{\mu,A})$, and $\hat{m}_{N,A}$ and $\hat{U}_{\mu,A}$ are the ML estimates for the HNL mass and coupling obtained using $\vec{d}_A$. The probability distribution, $f_t$, is obtained by generating $\mathcal{O}(10^3)$ pseudo-experiments, where the nuisance parameters have been fixed to their best-fit values.

Additionally, we can obtain parameter estimation contours by considering relative likelihoods using the test statistic
\begin{align}
    \label{eq: PLR}
    \tilde{t}(\vec{d} \,|\, m_N, U_\mu) := -2\ln \frac{L(\vec{d}\, | \, m_N , U_\mu;\,\hat{\eta},\hat{\vec{\gamma}}) }{L(\vec{d}\, | \, \hat{m}_N, \hat{U}_\mu;\,\hat{\eta},\hat{\vec{\gamma}})} \ .
\end{align}
CL contours can then be determined by computing the median $p$-value of this test statistic for each point in the model parameter space. Under the assumption that the underlying model is correct, Wilks' theorem states that the profiled likelihood ratio, \cref{eq: PLR}, should follow a $\chi^2$ distribution with degrees of freedom equal to the number of free parameters, which allows us to identify $1\sigma$, $2\sigma$, $3\sigma$, and $4\sigma$ CL thresholds at $\tilde{t}(\vec{d}_A) =  2.30$, 6.18, 11.83, and 23.94.

\subsection{Model Discrimination}

\begin{figure*}[tbph]
\centering
\includegraphics[width=0.49\linewidth]{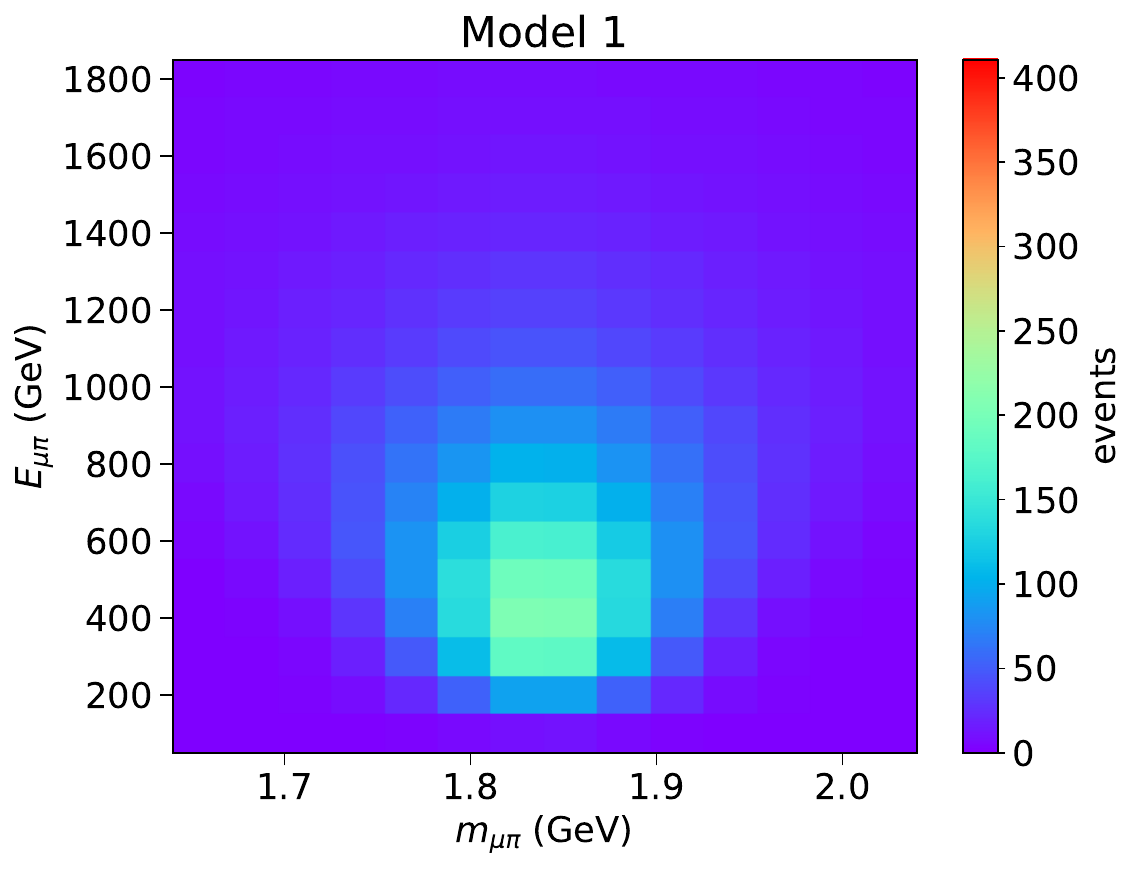}
\hfill
\includegraphics[width=0.49\linewidth]{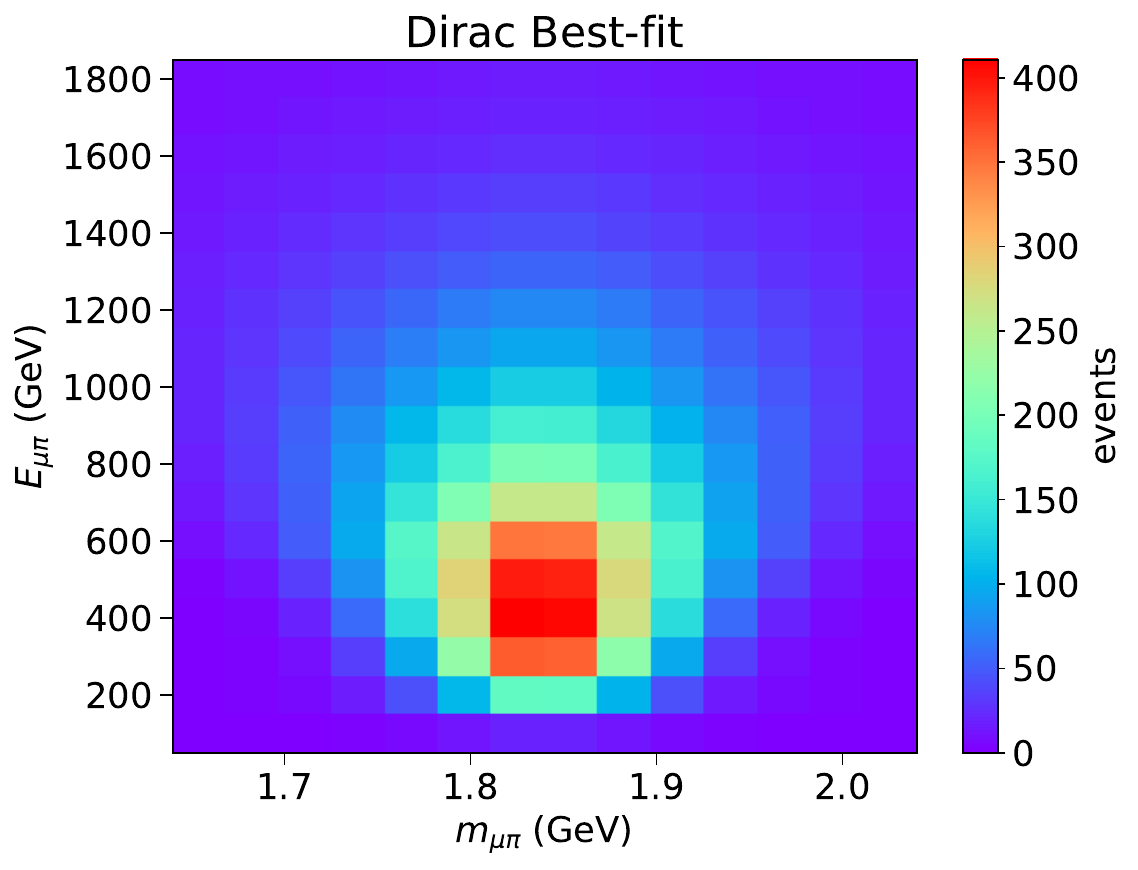}
\caption{Two-dimensional kinematic distributions for the Model 1 Asimov dataset, a Majorana HNL with $m_N = 1.84$ GeV and $U_\mu = 0.0036$ (left), and the best-fit Dirac model with $\hat{m}_N = 1.84$ GeV and $\hat{U}_\mu = 0.0051$, obtained by minimizing $t(\vec{d}_A\,|\,m_N,U_\mu)$ (right). The Dirac best-fit point recreates the signal shape well, but at the expense of a significantly larger event rate. }
\label{fig:2D FASER 1.8}
\end{figure*}

To differentiate between the Majorana and Dirac HNL models, we define the test statistic
\begin{align}
    \label{eq:lam}
    \lambda(\vec{d}\,) :=&-2\ln \frac{L_D(\vec{d}\, | \, \hat{m}_N^D ,\hat{ U}_\mu^D;\,\hat{\eta}^D,\hat{\vec{\gamma}}^D) }{L_M(\vec{d}\, | \, \hat{m}_N^M, \hat{U}_\mu^M;\,\hat{\eta}^M,\hat{\vec{\gamma}}^M)} \\ 
    =& \ t_D(\vec{d} \, | \, \hat{m}_N^D, \hat{U}_\mu^D)  - t_M (\vec{d} \, | \, \hat{m}_N^M , \hat{U}_\mu^M) \ ,
\end{align}
which is the likelihood ratio of the Majorana and Dirac models, which have different expected bin counts, for a given set of data $\vec{d}$. If $\lambda > 0$, the signal appears more ``Majorana-like,'' and if $\lambda < 0$, the signal appears more ``Dirac-like.'' We find the significance of this test by computing its median $p$-value,
\begin{align}
    \label{eqn:plam}
    p = \int^{\infty}_{\lambda_\text{exp}} f^D_{\lambda}(\lambda')\,d\lambda' \ ,
\end{align}
where $f^D_{\lambda}$ is the $\lambda$ distribution assuming the Dirac hypothesis is true, and $\lambda_\text{exp} = \lambda(\vec{d}_A)$. This represents the median probability that a Dirac HNL would fluctuate to produce a signal that looks just as (or more) ``Majorana-like'' than the current observation. Similar to $t$, we calculate the $p$-value by generating the probability distribution, $f_\lambda$, from $\mathcal{O}(10^3)$ pseudo-experiments, where the nuisance parameters for both models have been fixed to their best-fit values.

\section{Measuring HNL Parameters with FASER2}
\label{sec:Measuring model paremeters with FASER2}

\begin{figure*}[tbph]
\centering
\includegraphics[width=0.49\linewidth]{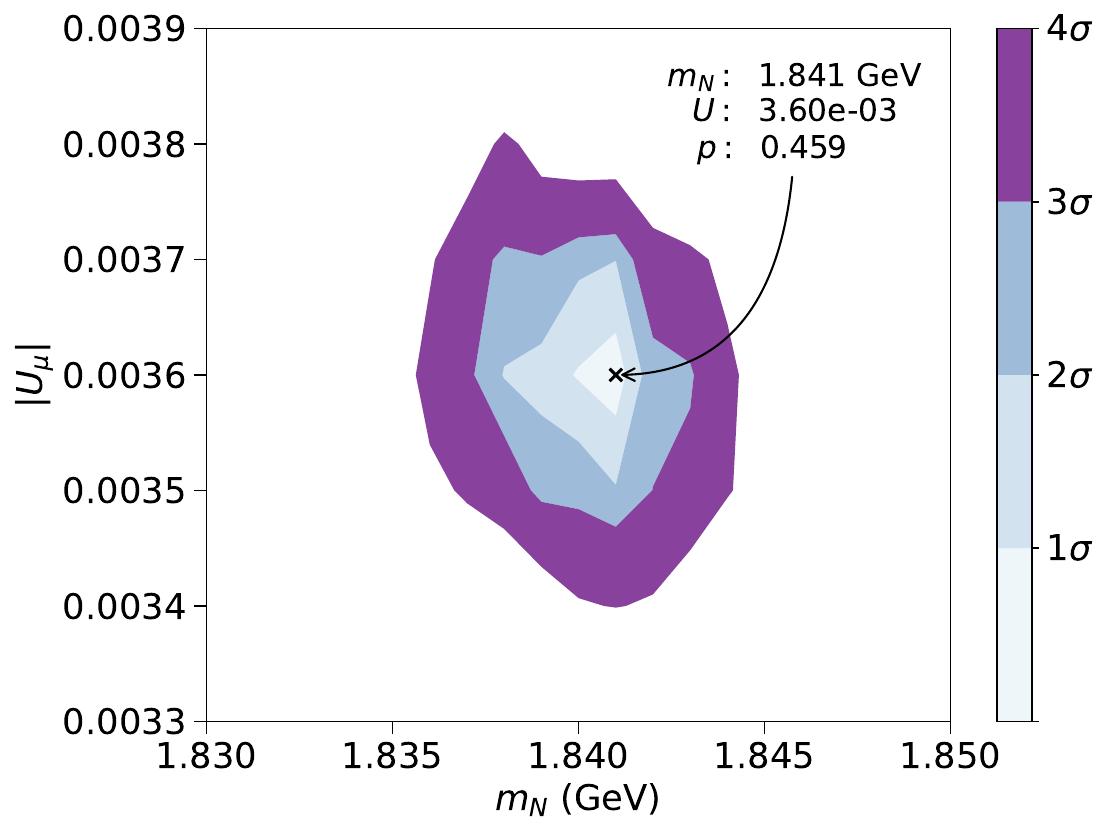}
\hfill
\includegraphics[width=0.49\linewidth]{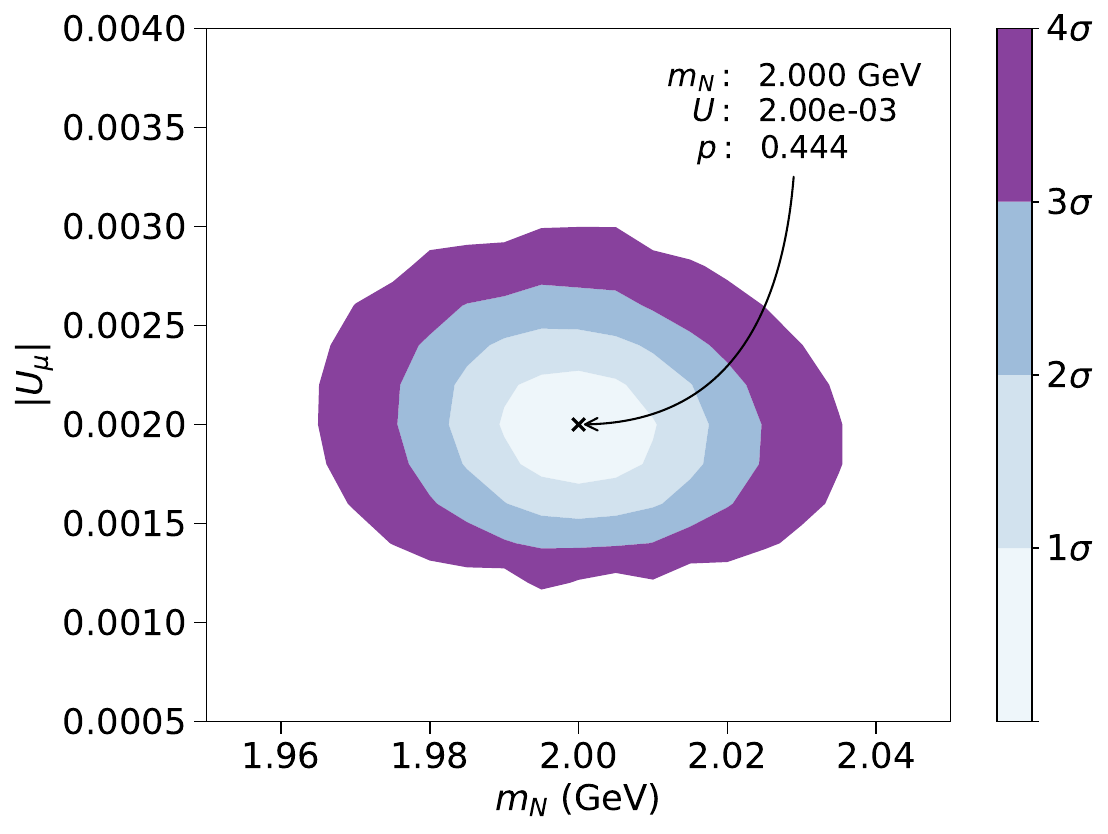}
\caption{Expected parameter estimation contours, assuming the underlying data corresponds to Model 1, a Majorana HNL with $m_N = 1.84~\text{GeV}$ and $U_{\mu} = 0.0036$ (left), and Model 2, a Majorana HNL with $m_N = 2.0~\text{GeV}$ and $U_{\mu} = 0.002$ (right). The HNL flux normalization and shape uncertainties are assumed to be $\sigma_{\eta} = 60\%$ and $\sigma_{\gamma} = 10\%$, respectively. }
    \label{fig:estimation 1.8}
\end{figure*}

\begin{figure*}[tbph]
    \centering
\includegraphics[width=0.48\linewidth]{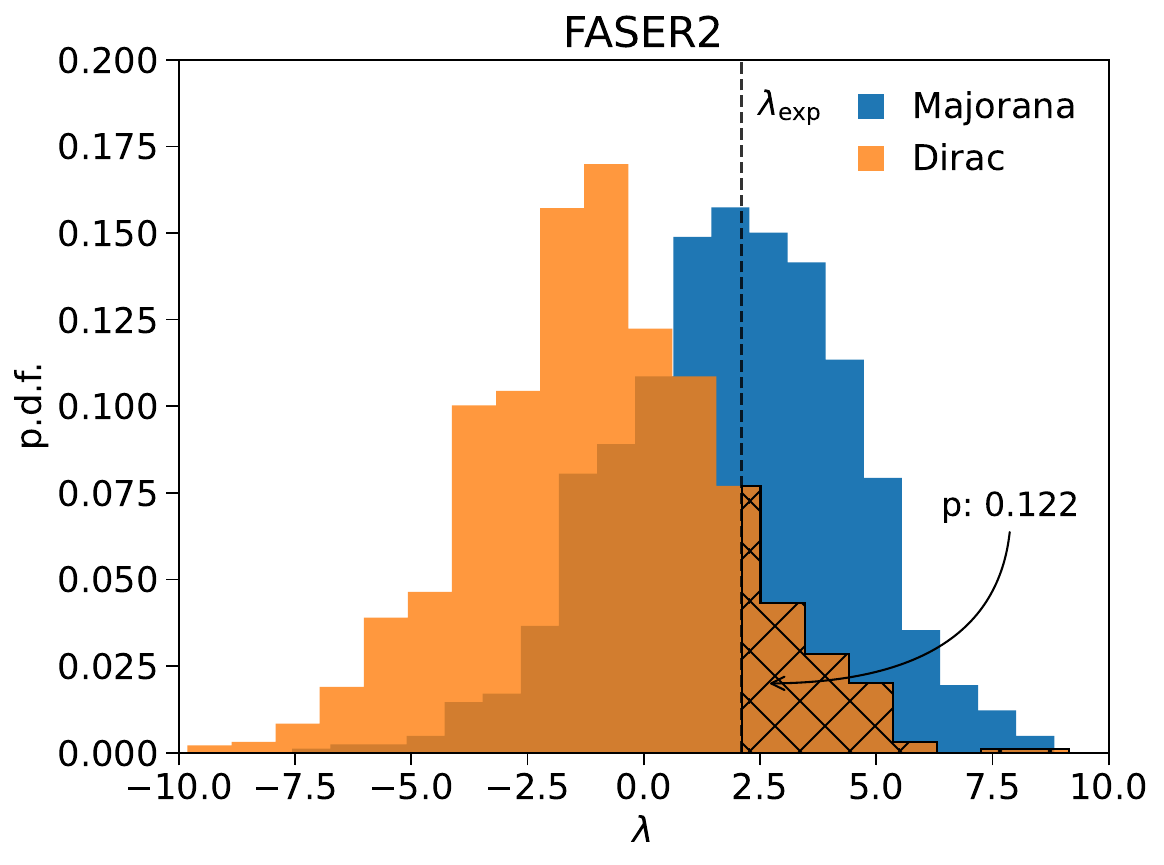}
    \hfill
\includegraphics[width=0.48\linewidth]{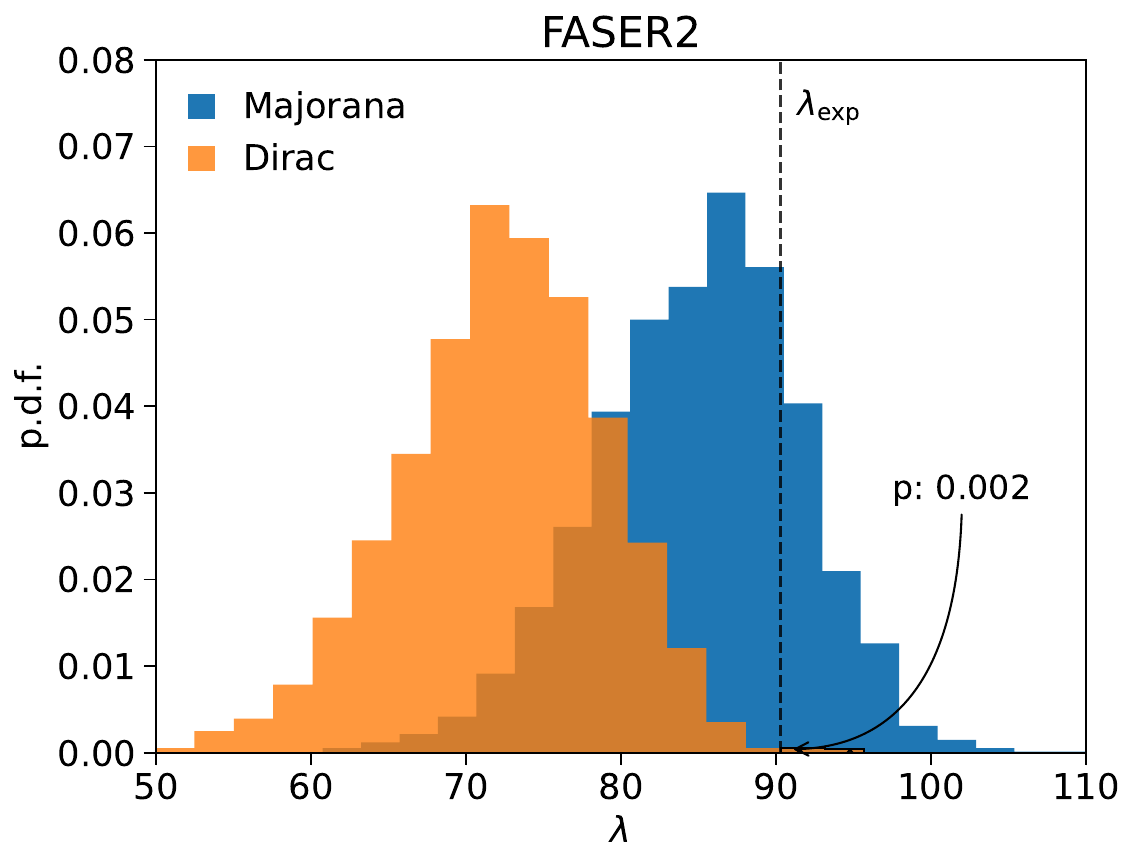}
\caption{Probability distributions of the likelihood ratio between the Majorana and Dirac models, $\lambda$, using FASER2 measurements under the Majorana and Dirac hypotheses, respectively. The underlying data is assumed to be Model 1, a Majorana HNL with $m_N = 1.84~\text{GeV}$ and $U_{\mu} = 0.0036$.  Left: Assuming HNL flux normalization and shape uncertainties of $\sigma_\eta = 60\%$ and $\sigma_\gamma = 10\%$, respectively, the Dirac hypothesis can be rejected at 87.8\% CL ($1.2\sigma$).  Right: Assuming a flux normalization uncertainty of $\sigma_\eta = 5\%$ and a negligible shape uncertainty $\sigma_\gamma$, the Dirac hypothesis can be rejected at 99.8\% CL ($2.9 \sigma$). }
    \label{fig:lambda FASER2 1.8}
\end{figure*}

\begin{figure*}[tbph]
    \centering
\includegraphics[width=0.48\linewidth]{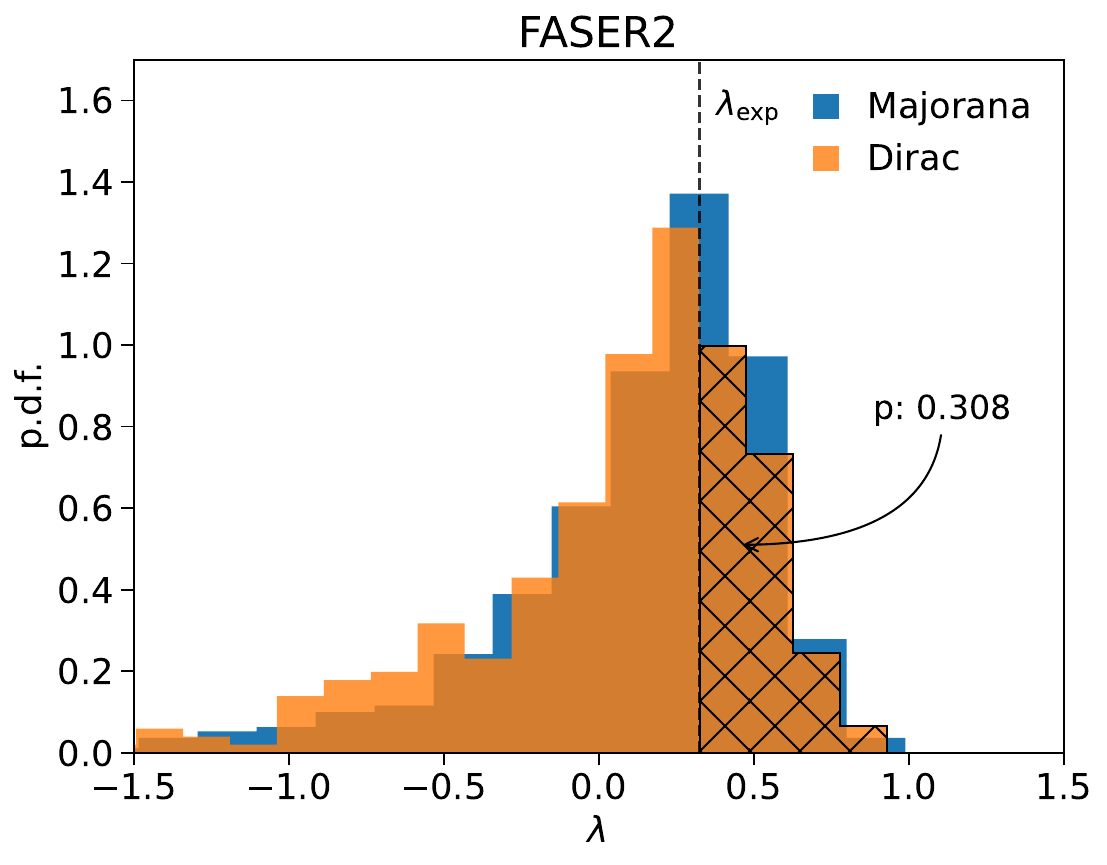}
    \hfill
\includegraphics[width=0.48\linewidth]{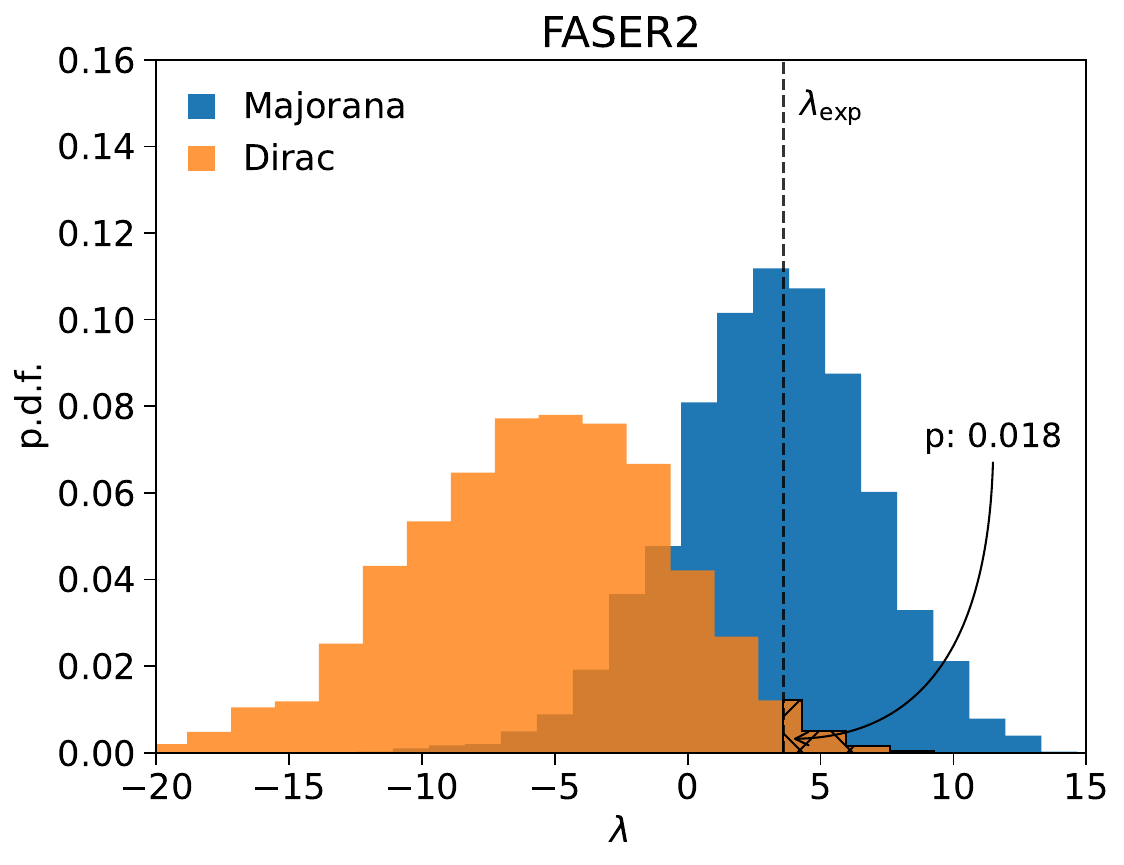}
\caption{Probability distributions of the likelihood ratio between the Majorana and Dirac models, $\lambda$, using FASER2 measurements under the Majorana and Dirac hypotheses, respectively. The underlying data is assumed to be Model 2, a Majorana HNL with $m_N = 2.0~\text{GeV}$ and $U_{\mu} = 0.002$. Left: Assuming HNL flux normalization and shape uncertainties of $\sigma_\eta = 60\%$ and $\sigma_\gamma = 10\%$, respectively, the Dirac hypothesis can be rejected at 69.2\% CL ($0.5\sigma$).  Right: Assuming a flux normalization uncertainty of $\sigma_\eta = 5\%$ and a negligible shape uncertainty $\sigma_\gamma$, the Dirac hypothesis can be rejected at 98.2\% CL ($2.1 \sigma$). }
\label{fig:lambda FASER2 2.0}
\end{figure*}

With the aim of constraining $(m_N, U_\mu)$, we separate simulated $N\to\mu\pi$ events into kinematic bins based on the invariant mass, $m_{\mu\pi}$, and total energy, $E_{\mu\pi}$, of the final state. The invariant mass of the final state is reconstructed via \begin{align}
        m_{\mu\pi}^2 = m_\mu^2 + m_\pi^2 + 2(E_\mu E_\pi-p_\mu p_\pi\cos\theta_{\mu\pi}) \ ,
    \end{align}
where $E_{\mu(\pi)}$ is the energy of the muon (pion), $p_{\mu(\pi)}$ is the 3-momentum of the muon (pion), and $\theta_{\mu\pi}$ is the opening angle between the muon and pion, all of which are smeared according to the resolutions given in \cref{sec:Simulating HNL signals at FASER2}. 

To understand the measurement capabilities of FASER2, we conduct our analysis on an Asimov dataset representing Model 1, a Majorana HNL with $m_N = 1.84$~GeV and $U_\mu = 0.0036$, which predicts a large number of $N\to\mu\pi$ events in the FASER2 detector.  The resulting 2D kinematic distribution of the Model 1 dataset is shown in the left panel of \cref{fig:2D FASER 1.8}.

In the left panel of \cref{fig:estimation 1.8}, we present results showing the expected FASER2 constraints on $(m_N,U_\mu)$ in the Majorana parameter space under the assumption that Model 1 is true. We find that FASER2 has the ability to measure $m_N$ to about 2~MeV (a fractional uncertainty of $0.1\%$) and $U_\mu$ with a fractional error of approximately $3\%$ at $95\%$ CL. Additionally using \cref{eq:p good}, we find the expected goodness-of-fit $p$-value to be $0.46$ ($0.1\sigma$), suggesting consistency with the Asimov dataset, as expected.

\section{Differentiating Majorana versus Dirac HNLs at FASER2}
\label{sec:Discrimination on the basis of lifetime}

Using the test statistic $\lambda$ defined in \cref{eq:lam}, we differentiate between the Majorana and Dirac models by generating the distribution of $\lambda$ under both assumptions and comparing their distributions to the observed value $\lambda_{\text{exp}}$ using \cref{eqn:plam}.  We present the resulting distributions after 1000 pseudo-experiments in \cref{fig:lambda FASER2 1.8}. We find that with HNL flux normalization and shape uncertainties of $\sigma_\eta = 60\%$ and $\sigma_\gamma = 10\%$, respectively, we expect that FASER2 would only favor the Majorana model over the Dirac model at 87.8\% CL ($1.2\sigma$), as shown in the left panel of \cref{fig:lambda FASER2 1.8}.

To understand this surprising result, in the right panel of \cref{fig:2D FASER 1.8}, we show that the Dirac model best-fit recreates the shape of the mass and energy distributions of the data very well, but overestimates the total event rate by a factor of 2. However, this falls within the large normalization uncertainty in the HNL flux, resulting in limited discrimination capability. 

However, as FASER$\nu$ continues to constrain hadron fluxes from neutrino observations~\cite{FASER:2024ref}, the uncertainty in hadron production from the MC generators will be reduced significantly by data. These constraints may be sharpened further by additional data from FASER$\nu$2 and FLArE at the Forward Physics Facility~\cite{Adhikary:2024nlv,FPFWorkingGroups:2025rsc}. If the normalization uncertainty is reduced to $\sigma_\eta = 5\%$ and the shape uncertainty $\sigma_\gamma$ becomes negligible, this would exacerbate the difference in event rates and increase our expected discrimination capability to 99.8\% CL ($2.9 \sigma$), as shown in the right panel of \cref{fig:lambda FASER2 1.8}.

\section{FASER2 as a Trigger for ATLAS}
\label{sec:FASER2 as a trigger for ATLAS}

So far, we have considered what can be learned with FASER2 measurements alone.  In optimistic scenarios where there are many events, as in Model 1 of \cref{tab:model_summary}, which we have considered so far, we have seen that FASER2 may precisely measure the HNL parameters, and help determine if it is a Majorana or Dirac fermion if hadronic fluxes are better constrained.

As noted in \cref{sec:intro}, however, it is also possible to discover HNLs with far fewer events, and it is interesting to explore what can be done in this case. In this section, we consider Model 2 of \cref{tab:model_summary}, where only approximately 80 HNL decays $N \to \mu \pi$ can be seen at FASER2 at the HL-LHC. 

Applying the same method and statistical analysis as done for Model 1, we find that for Model 2, FASER2 alone can still constrain the HNL mass and coupling to fractional uncertainties of $1\%$ and $25\%$ at $95\%$ CL, respectively, as shown in the right panel of \cref{fig:estimation 1.8}. The mass measurement is still quite precise, but the coupling measurement is significantly degraded relatively to Model 1. Additionally, we find that the Dirac hypothesis can be disfavored at 98.2\% CL ($2.1 \sigma$) with improved hadronic fluxes, as shown in \cref{fig:lambda FASER2 2.0}. The ability to distinguish Majorana from Dirac is therefore slightly degraded relative to the case of Model 1.

In the rest of this section, we discuss whether FASER2 may be used as a trigger for ATLAS, allowing ATLAS to record muons produced in association with the HNL, and we discuss what improvements to model discrimination can be made by the combining measurements at the two detectors.

\subsection{Forward Muons at ATLAS}

The HL-LHC is scheduled to begin operating in 2030 and continue through to the planned end of the LHC in the 2040s. During this period, the peak instantaneous luminosity is expected to rise from its current value of $1 \times 10^{34} \, \text{cm}^{-2} \, \text{s}^{-1}$, producing on average 27 pileup events per bunch crossing, to $7.18 \times 10^{34} \, \text{cm}^{-2} \, \text{s}^{-1}$, producing around 200 pileup events per bunch crossing. The goal is to accumulate an integrated luminosity of approximately $3~\text{ab}^{-1}$ by the end of the HL-LHC era~\cite{Apollinari:2015PreliminaryDesignHL-LHC, ZurbanoFernandez:2020cco}.

In conjunction with the HL-LHC upgrade, the ATLAS detector will undergo a significant Phase-II upgrade. This includes the installation of a completely new Inner Tracker (ITk), which will replace the existing Pixel Detector, Semiconductor Tracker (SCT), and Transition Radiation Tracker (TRT). Additional upgrades include major improvements to the Trigger and Data Acquisition (TDAQ) system and discussion of an additional high $\eta$ muon tagger. The muon tagger will extend muon coverage to regions up to $|\eta| < 4$, significantly increasing overall acceptance~\cite{CERN-LHCC-2015-020, ATLAS:2009zsq, Henriques:2015TileCal}.

Calorimeter systems at ATLAS also play a critical role in particle identification and energy measurement. The electromagnetic calorimeter provides coverage up to $|\eta| < 3.2$~\cite{ATLAS:2009zsq}, while the Tile Hadronic Calorimeter covers the central region up to $|\eta| < 1.7$~\cite{ATLAS:2024ffc, Henriques:2015TileCal}. The Hadronic Endcap Calorimeter (HEC) extends coverage into the $1.5 < |\eta| < 3.2$ range~\cite{ATL-LARG-PUB-2022-001}. Complementing these, the Forward Calorimeter (FCal) comprises three modules, FCal1 (EM) and FCal2/FCal3 (hadronic) which covers the very forward region from $3.1 < |\eta| < 4.9$~\cite{Gingrich:2007ia, Henriques:2015}.

Taken together, the tracking and calorimeter systems provide full coverage for energy measurements and charged particle reconstruction up to $|\eta| < 4.9$, with muon identification capabilities extending to $|\eta| < 4$. For this analysis, we consider an optimistic scenario in which both the inner tracker and muon tagger are extended to cover $|\eta| < 4.9$. Such an extension would enhance the prospects for identifying muons and their charge in the forward region.

An intriguing possibility is that FASER2 could be used as a trigger for ATLAS. FASER2 is capable of issuing a trigger signal within approximately 1~ns~\cite{FASER:2022hcn}. As noted in \cref{sec:intro}, the total time for a round trip signal, accounting for both the time it takes the HNL to propagate to FASER2 at the speed of light and the return signal through an optical fiber back to ATLAS, is approximately 5 $\mu$s. The ATLAS L0 trigger latency is 6 $\mu$s~\cite{CERN-LHCC-2015-020}. Therefore, it is feasible to instruct ATLAS to retain data from a bunch crossing that has produced an HNL signal in FASER2. This coordination would allow for the simultaneous use of FASER2 and ATLAS data to perform a correlated analysis~\cite{Pastore:2024, 2137107, 2106380}.

\subsection{Pileup Background}

\begin{figure}[tbp]
\centering
\includegraphics[scale=0.5]{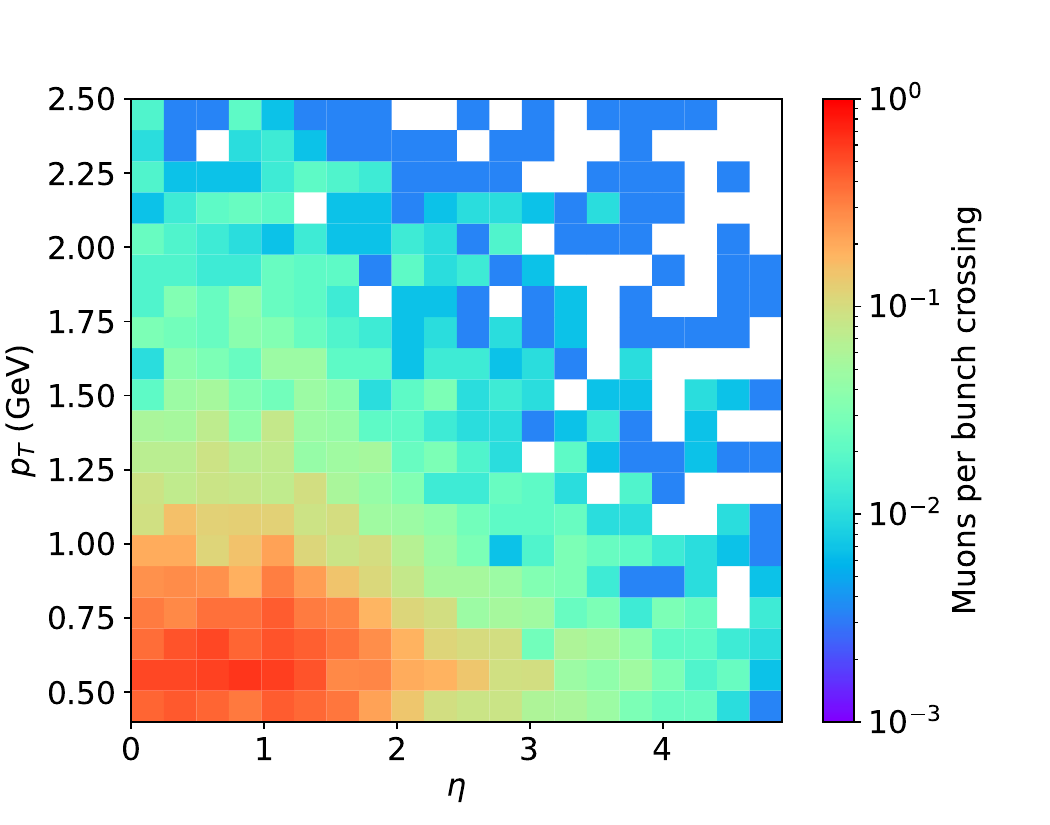}
\caption{The number of muons produced per bunch crossing in the $(\eta, p_T)$ plane, including both muons that are produced in light meson ($\pi^{\pm}$, $K^{\pm}$, $K^0_L$) decays and heavy meson (charm and bottom meson) decays within the ITk.}
    \label{fig:bkg_pt_eta}
\end{figure}

The muons produced in association with HNLs at ATLAS are produced in the forward region, where there is a lot of background, especially given the effects of pileup.  To estimate the pileup background, we simulate proton-proton collisions at a center-of-mass energy of 14 TeV using \texttt{Pythia8}.  We conservatively assume the peak instantaneous luminosity, corresponding to roughly $200$ pileup events per bunch crossing, which gives the maximal expected background.  The simulation includes all muons originating from either light or heavy mesons that decay within the ITk, which is modeled as a cylindrical volume with a radius of 1~m and a length of 6~m.  The SoftQCD option in \texttt{Pythia8} is turned on to simulate low-energy QCD interactions. 

\begin{table}[tbp]
\centering
\begin{tabular}{|c|c|c|}
\hline
Parent  & Muons ($|\eta|<4.9$) & Muons ($3.5 < \eta <4.9$) \\
\hline
$\pi^{\pm}$    & 31.41 & 0.43  \\
$K^{\pm}$    & 16.09 & 0.26 \\
$K^0_L$    & 0.37 & 0.0033 \\
Heavy Mesons & 3.34 & 0.33  \\
\hline
\end{tabular}
\caption{Number of muons produced with $p_T > 0.4$ GeV in the ITk per bunch crossing at ATLAS from various parent particles, shown for the full muon coverage $|\eta| < 4.9$ and the forward region $3.5 < \eta < 4.9$. ``Heavy Mesons'' includes charm and bottom mesons. Yields include both $\mu^+$ and $\mu^-$. Muons are categorized by their direct parent particle, so, for example, muons from $K^0_S \to \pi^+ \pi^-$, with the pions decay to muons, are included in the $\pi^{\pm}$ row. The number of muons produced directly from $K^0_S$ decays is negligible. Muons from light meson decays dominate the background, yielding around 50 per bunch crossing in the full acceptance, but only about 0.7 within the forward region after cuts. }
\label{tab:summary_final}
\end{table}

The simulation assumes a muon detector with coverage up to $|\eta| < 4.9$, as well as a shield that allows one to distinguish muons from charged pions. The resulting background-muon distribution in $(\eta, p_T)$ is shown in~\cref{fig:bkg_pt_eta}, peaking at lower $\eta$ values. In contrast, signal muons from HNL decays in FASER2 populate higher $\eta$, motivating the cut $3.5 < \eta < 4.9$ to retain most of the signal while rejecting the bulk of the background. An additional requirement of $p_T > 0.4$ GeV avoids the region of poor reconstruction efficiency. The surviving background rates are summarized in~\cref{tab:summary_final}.

Combining muon contributions from both light and heavy meson decays, we estimate that roughly 1 background muon per bunch crossing satisfies the final selection criteria. This sets a baseline for understanding and mitigating pileup backgrounds in the forward region of the detector, which is essential for assessing the feasibility of using FASER2 as a trigger for ATLAS.

\subsection{Results}

\begin{figure*}[tbph]
\centering
\includegraphics[width=0.49\linewidth]{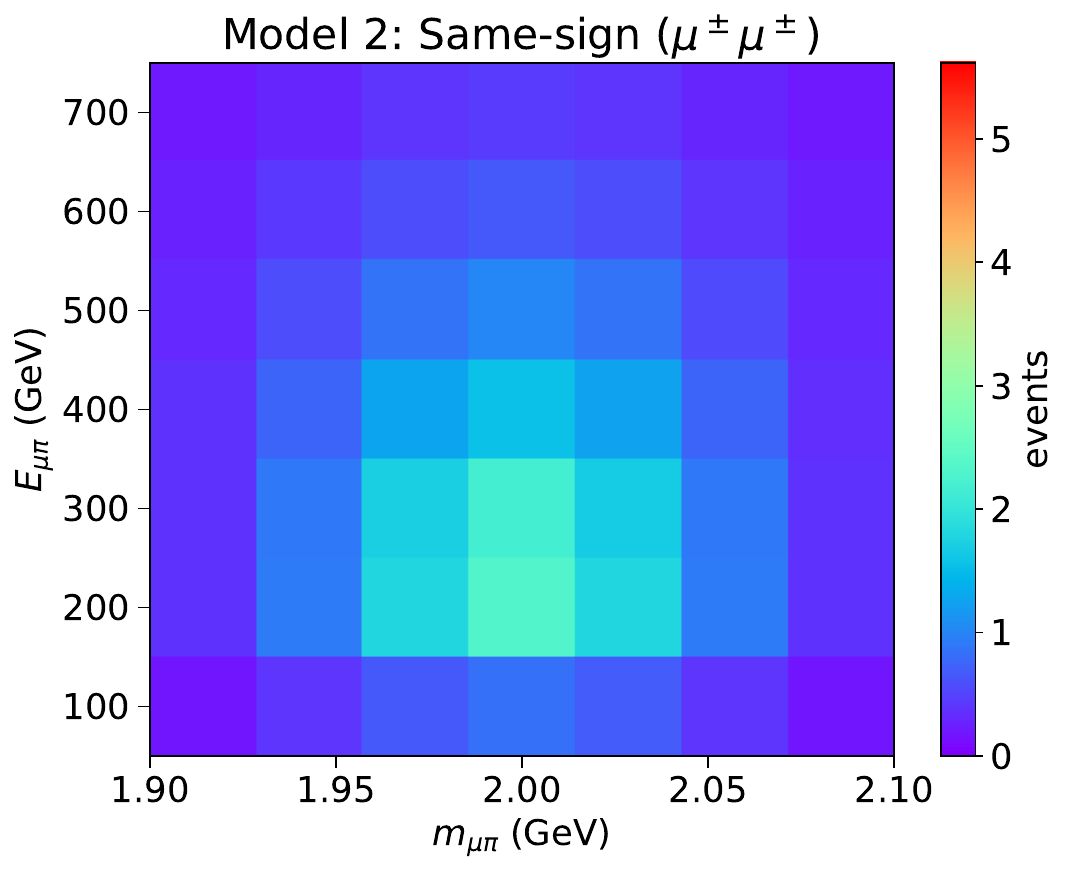}
\hfill
\includegraphics[width=0.49\linewidth]{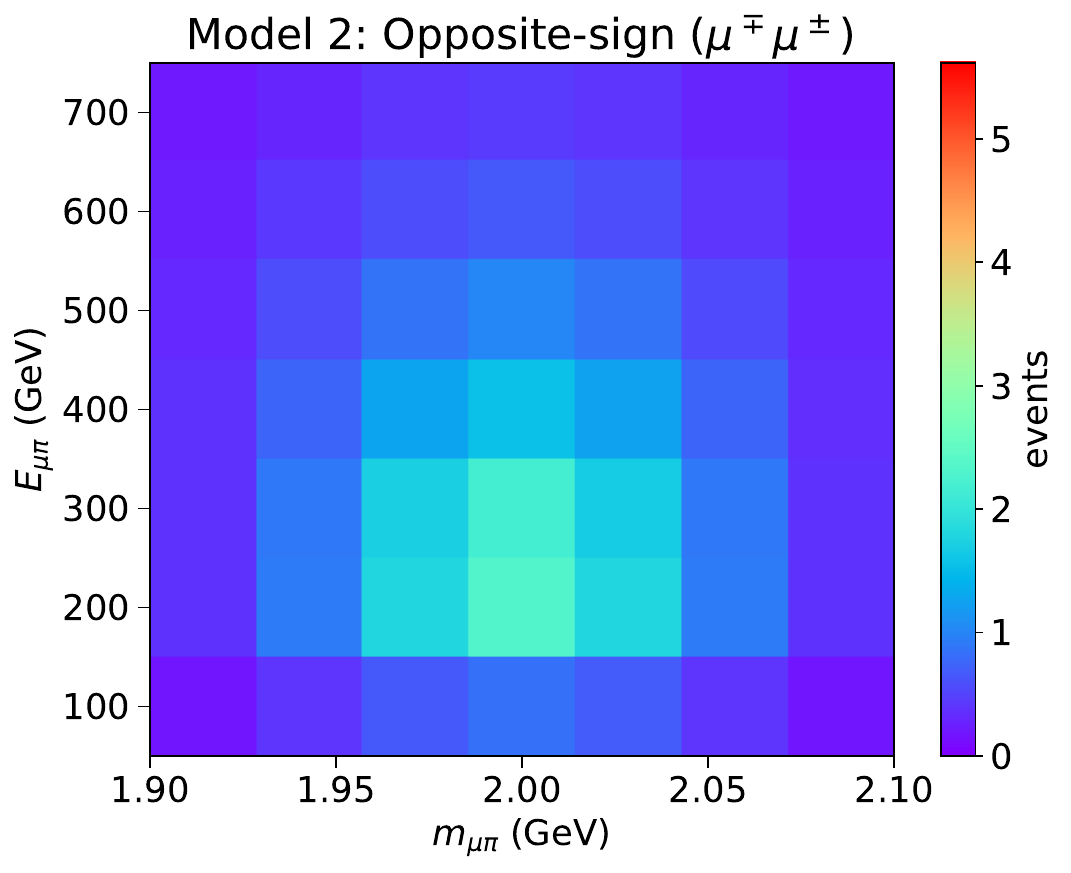}
\caption{Two-dimensional kinematic distributions for the Model 2 Asimov dataset with a Majorana HNL with $m_N = 2~\gev$ and $U_{\mu} = 0.002$ with SS (left) and OS (right) muons seen in ATLAS, assuming an integrated luminosity of $3~\text{ab}^{-1}$. These distributions include both muons from HNL events and background muons from pileup.  Muons produced in association with Majorana HNLs can be found with OS or SS muons in FASER2 with equal rates, resulting in a symmetric distribution between the two histograms. }
    \label{fig:2D ATLAS Signal}
\end{figure*}

\begin{figure*}[tbph]
    \centering
\includegraphics[width=0.49\linewidth]{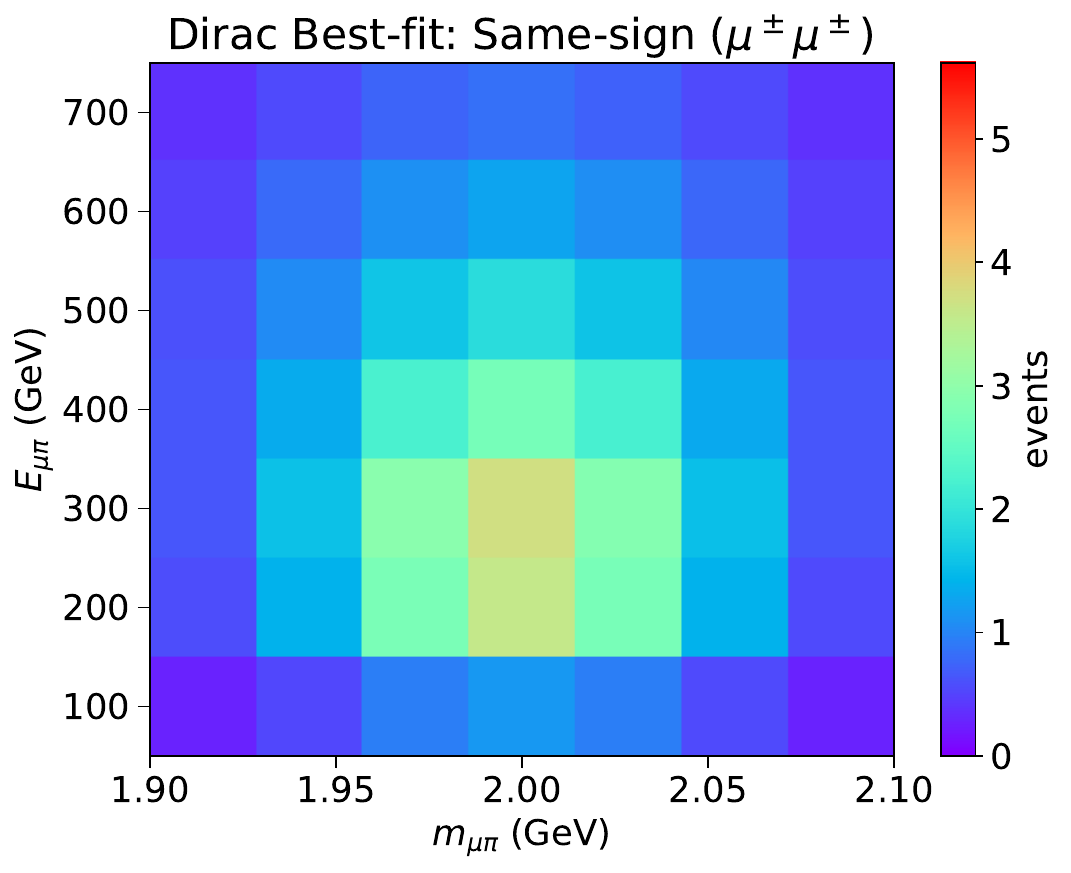}
\hfill
\includegraphics[width=0.49\linewidth]{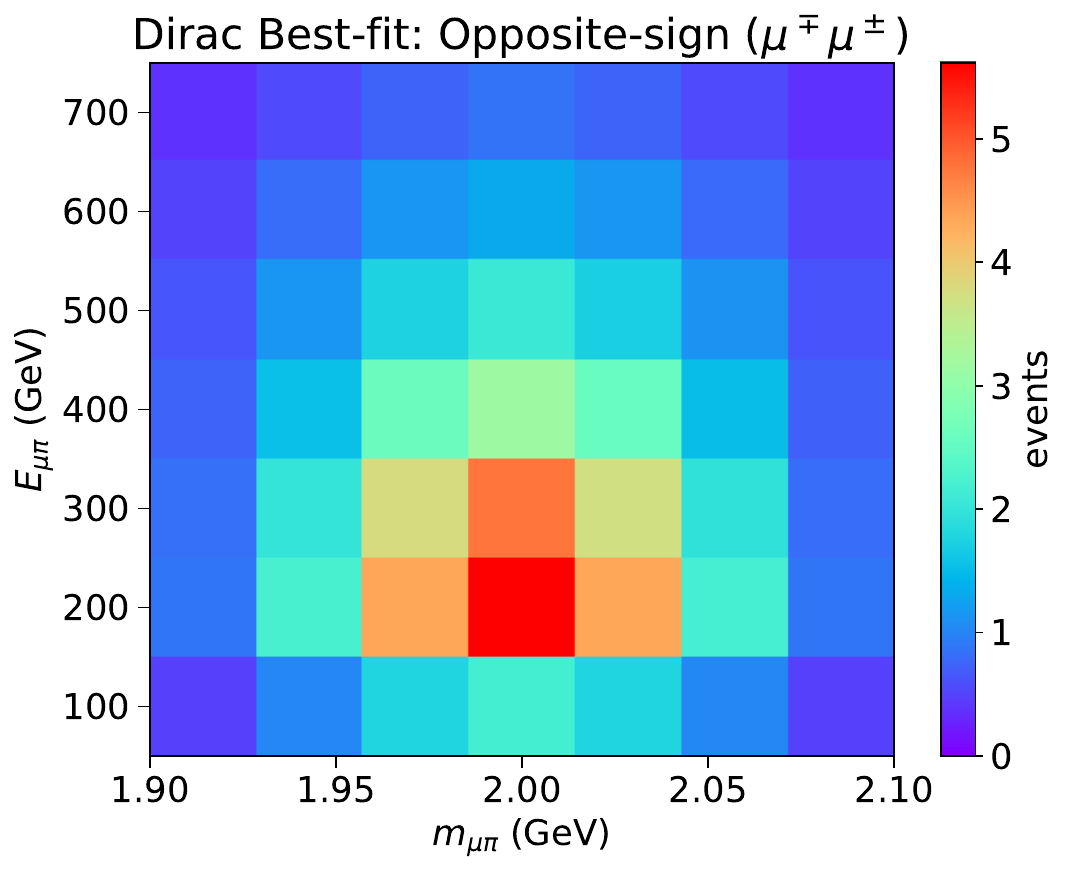}
\caption{Two-dimensional kinematic distributions for the Dirac HNL best-fit point to the Model 2 Asimov dataset with SS (left) and OS (right) muons seen in ATLAS, obtained by minimizing the test-statistic $t(\vec{d}_A\,|\,m_N,U_\mu)$, assuming an integrated luminosity of $3~\text{ab}^{-1}$. These distributions include both muons from HNL events and background muons from pileup. Muons produced in association with Dirac HNL's can only be found with OS muons in FASER2, resulting in an asymmetric distribution between the two histograms. }
    \label{fig:2D ATLAS BF}
\end{figure*}

\begin{figure*}
    \centering
\includegraphics[width=0.49\linewidth]{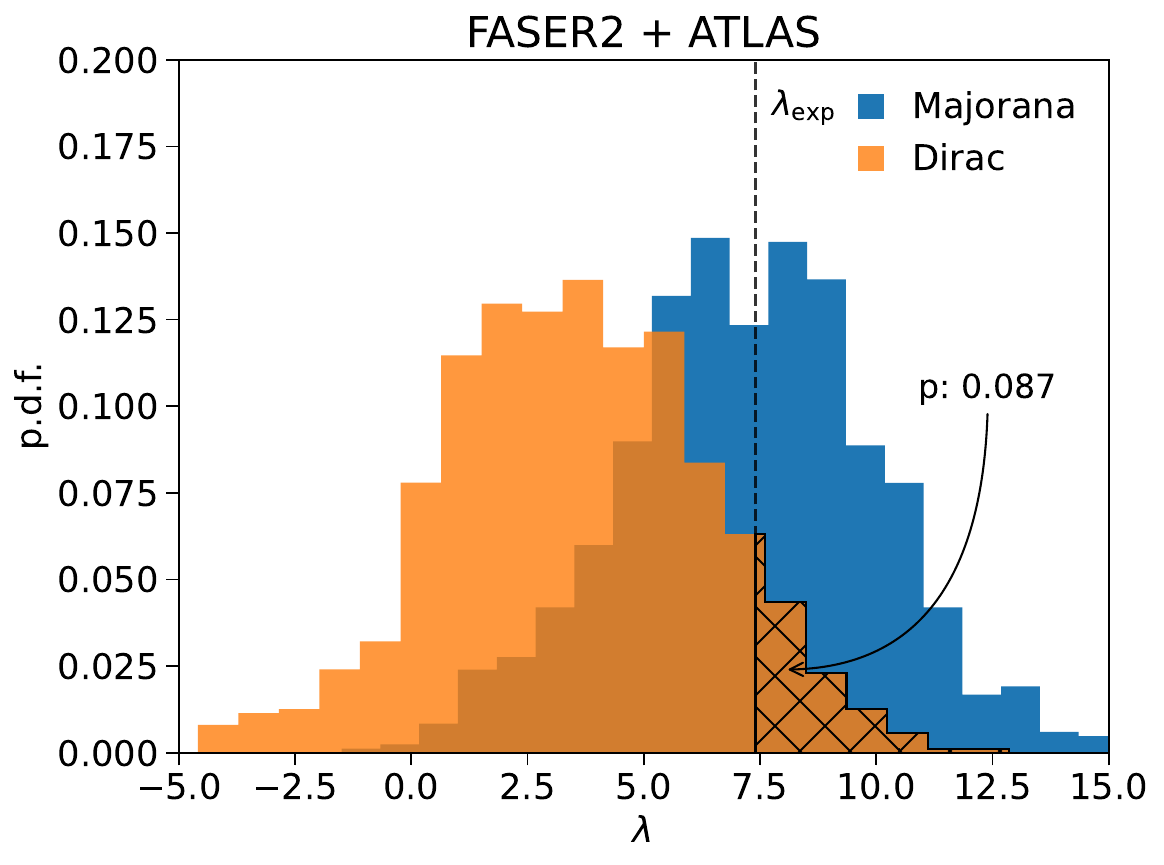}
\hfill  \includegraphics[width=0.49\linewidth]{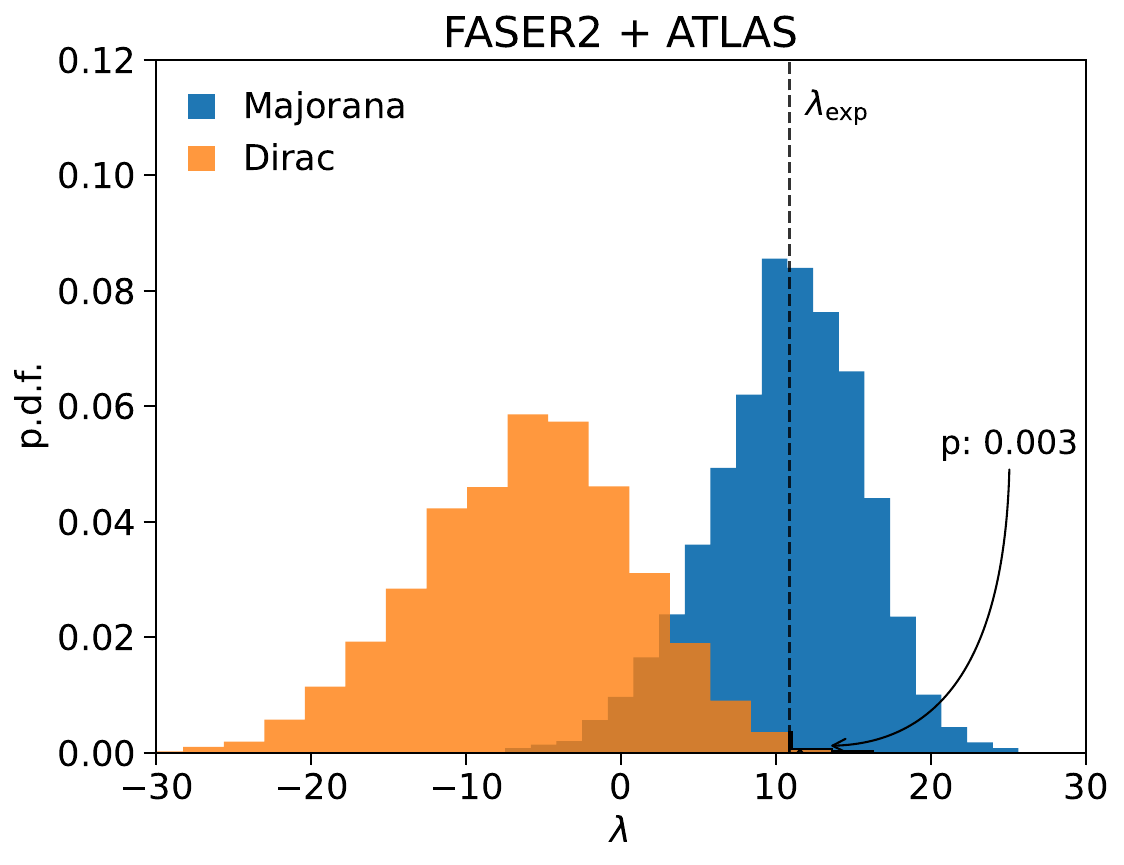}
\caption{Probability distributions of the likelihood ratio between the Majorana and Dirac models, $\lambda$, using correlated FASER2 and ATLAS data, under the Majorana and Dirac hypotheses, respectively. The underlying data is assumed to be Model 2, a Majorana HNL with $m_N = 2.0~\text{GeV}$ and $U_{\mu} = 0.002$.  Left: Assuming HNL flux normalization and shape uncertainties of 60\% and 10\%, respectively, the Dirac hypothesis can be rejected at 91.3\% CL ($1.4\sigma$).  Right: Assuming a flux normalization uncertainty of $\sigma_\eta = 5\%$ and a negligible shape uncertainty $\sigma_\gamma$, the Dirac hypothesis can be rejected at 99.7\% CL ($2.7 \sigma$). } 
    \label{fig:lambda ATLAS}
\end{figure*}

We separate each triggered HNL event into the number of muons detected in ATLAS that have either the same sign (SS) or opposite sign (OS) as the muon detected in FASER2.  Majorana HNLs decay to $\mu^+\pi^-$ and $\mu^-\pi^+$ and produce SS and OS signals with equal rates, whereas Dirac HNLs only produce OS signals. In the absence of background, a single SS event would be evidence of LNV and a Majorana HNL.  However, uncorrelated background muons produced in ATLAS may give ``fake'' SS signals, and so LNV and Majorana-ness can only be established on a statistical basis. Therefore, the expected number of SS and OS events in each bin is given by 
\begin{align}
    \mu^\text{SS}_i &= \frac{b}{2}\, \mu^{\text{F}}_i+ \frac{\eta_M}{2} \,\mu^\text{A}_i \\
    \mu^\text{OS}_i &= \frac{b}{2}\, \mu^{\text{F}}_i + \left(1-\frac{\eta_M}{2} \right) \,\mu^\text{A}_i  \ ,
\end{align}
where $\mu^\text{F}_i$ is the total expected number of HNL events in bin $i$ as measured in FASER2, $b \approx 1$ is the expected number of uncorrelated background muons per bunch crossing produced by pileup, $\mu^A_i$ is the expected number of truly correlated muons detected in ATLAS in bin $i$, and $\eta_M = 1$ for the Majorana model and $\eta_M = 0$ for the Dirac. The uncertainty in each bin is estimated by adding the errors of $\mu^{\text{F}}_i$ and $\mu_i^\text{A}$ in quadrature.  In \cref{fig:2D ATLAS Signal,fig:2D ATLAS BF}, we plot the two-dimensional kinematic distribution of the Model 2 Asimov dataset, and the Dirac best-fit point, respectively, separated into SS and OS bins where a clear asymmetry can be observed between the distribution of events that appear LNC and LNV in the Dirac case. In \cref{fig:lambda ATLAS}, we present results for the distribution of $\lambda$ utilizing the correlated events from ATLAS. We expect that the Dirac model would be rejected at 91.3\% CL ($1.4\sigma$), assuming a normalization uncertainty of $\sigma_\eta = 60\%$ and a shape uncertainty of $\sigma_\gamma = 10\%$. If the normalization uncertainty is reduced to $\sigma_\eta = 5\%$ and the shape uncertainty $\sigma_\gamma$ becomes negligible by improved constraints on forward hadron production, we find a major improvement in discrimination capability, and we expect to be able to reject the Dirac model at 99.7\% CL ($2.7 \sigma$). 

\section{Conclusions}
\label{sec:Conclusion}

In this work, we have investigated the potential of FASER2 to measure the mass and coupling of an HNL and determine whether it is Majorana or Dirac in the event that HNL decays $N\to\mu\pi$ are detected. Additionally, we have explored the possibility of using a signal in FASER2 to trigger ATLAS, and using the correlated FASER2 and ATLAS data to improve model discrimination by correlating the charge of the muon produced in association with the HNL at ATLAS with the charge of the muon produced in HNL decay at FASER2.  

When investigating the capabilities of FASER2 alone, we considered benchmark Model 1, a Majorana HNL with $m_N = 1.84$ GeV and $U_\mu = 0.0036$, which predicts approximately 8600 $N\to\mu\pi$ decays in FASER2 at the HL-LHC. We performed a 2D binned ML fit of the HNL model parameters to the expected dataset with events classified by the invariant mass and total energy of the $\mu \pi$ system. Assuming the currently large uncertainties in the forward hadron flux, with an overall flux uncertainty of $\sigma_{\eta} = 60\%$ and shape uncertainty of $\sigma_{\gamma} = 10\%$, we have found that the HNL's mass and couplings can be measured remarkably precisely to fractional uncertainties of approximately 0.1\% and 3\% at 95\% CL, respectively, but FASER2 would only be able to favor the Majorana model over the Dirac model at 87.8\% CL ($1.2\sigma$). However, assuming that far-forward neutrino measurements are able to reduce forward hadron flux uncertainties to $\sigma_{\eta} = 5\%$ and a negligible $\sigma_{\gamma}$, the Dirac fermion hypothesis can be rejected at 99.8\% CL ($2.9 \sigma$).

In the regions of parameter space where the signal yield is insufficient to differentiate Majorana and Dirac HNLs with FASER2 data alone, we have also explored the possibility of using FASER2 to trigger ATLAS and using the correlated FASER2 and ATLAS events to improve model discrimination. In our work, we considered benchmark Model 2, a Majorana HNL with $m_N = 2.0$ GeV and $U_\mu = 0.002$ which predicts only 80 $N\to \mu\pi$ events at FASER2 at the HL-LHC.  We determined the expected $\mu$ yield at ATLAS from HNL production, as well as the expected background from pileup interactions. Separating HNL events into those that appear LNC (with an OS muon observed in ATLAS) and those that appear LNV (with an SS muon observed in ATLAS), we found that FASER2 and ATLAS data combined were able to discriminate between Majorana and Dirac HNLs, even in the presence of misidentified background events (\cref{fig:2D ATLAS Signal,fig:2D ATLAS BF}). We found that if the normalization uncertainty from forward hadron production was reduced to $\sigma_{\eta} = 5\%$ and the shape uncertainty $\sigma_\gamma$ was negligible, the combined data from FASER2 and ATLAS are able to reject the Dirac hypothesis at 99.7\% CL ($2.7 \sigma$). 

Further work could extend this analysis by consider more general HNL models as well as other HNL decays. For the sake of simplicity, we limited ourselves to the case of an HNL with only muon couplings.  However, models with couplings to more than one flavor have been suggested as being more representative of realistic models that generate the observed neutrino masses and mixing angles~\cite{Drewes:2022akb} and thus pose an interesting avenue for further study. Additionally, we chose to focus on the decay channel $N\to \mu\pi$, where the decay products are fully visible and the $N$ energy and momentum are therefore, in principle, fully reconstructible. However, another HNL decay mode, $\mu\rho$, although more complicated, also satisfies this criterion, and other decay modes, such as $\nu\mu\mu$, could be incorporated to further increase the overall signal yield and provide further characterization capability. 

This work shows that the main LHC detectors and proposed auxiliary detectors can have a synergistic relation on an event-by-event basis. In Model 2, FASER2 would detect HNL events, but ATLAS alone would not. However, by using FASER2 as a trigger for ATLAS, the combined data from FASER2 and ATLAS are able to detect the HNL over its entire lifetime from the cradle to the grave, that is, from its production to its decay.   With this combined information, the combination of transverse and far-forward LHC detectors may be able to determine the HNL's mass and couplings and also establish that the HNL is a Majorana fermion. Such studies would have far-reaching implications for many topics, including neutrino masses, neutrinoless double beta decay experiments, baryogenesis, and our understanding of the underlying symmetries of the physical world.

\section*{Acknowledgments}

We are grateful to Alan Barr and Iacopo Vivarelli for providing projections for FASER2, Anyes Taffard for helpful information about ATLAS, Felix Kling for insights regarding forward hadron production, and Takeo Moroi for a very helpful conversation. This work is supported in part by U.S.~National Science Foundation (NSF) Grants PHY-2210283 and PHY-2514888, Simons Foundation Grant \#623683, and Simons Investigator Award \#376204.  The work of J.L.F.~is supported in part by NSF Grant PHY-2111427 and Heising-Simons Foundation Grants 2019-1179 and 2020-1840.  D.W.~is supported by the DOE Office of Science. 

\bibliography{hnls}

\end{document}